\documentclass[pre,aps,floats,superscriptaddress,floatfix]{revtex4}
\usepackage{amssymb,amsmath}
\usepackage{amsmath,amssymb}
\usepackage{graphicx}
\usepackage{psfrag}
\usepackage{multirow}

\def\beq{\begin{equation}}
\def\eeq{\end{equation}}
\def\bea{\begin{eqnarray}}
\def\eea{\end{eqnarray}}
\begin{document}
\title{Role of interfacial friction for flow instabilities in a thin polar ordered
active fluid layer}
\author{Niladri Sarkar}\email{nsarkar@pks.mpg.de}
\author{Abhik Basu}\email{abhik.basu@saha.ac.in}
\affiliation{Condensed Matter Physics Division, Saha Institute of
Nuclear Physics, Calcutta 700064, India}
\affiliation{Max-Planck Institut f\"ur Physik Komplexer Systeme, N\"othnitzer Str. 38,
Dresden, D-01187 Germany}

\date{\today}
\begin{abstract}
We construct a generic coarse-grained dynamics of a thin
inflexible planar layer of polar-ordered suspension of active particles, that is
frictionally coupled to an
embedding isotropic passive fluid medium with a friction coefficient $\Gamma$.
Being controlled by $\Gamma$, our model provides a unified framework to
describe the long wavelength behaviour
of a variety
of thin polar-ordered systems, ranging from {\em wet} to {\em dry}  active
matters and free standing  active films.
Investigations of the
linear instabilities around a chosen orientationally ordered uniform
reference state reveal generic moving and static instabilities in
the system, that can depend sensitively on $\Gamma$. Based on our
results, we discuss
estimation of bounds on $\Gamma$ in experimentally accessible systems.

\end{abstract}

\maketitle

\section{Introduction}

The emergence of large-scale collective dynamics is one of the most
intriguing and fascinating features of a large variety of driven,
active systems made of {\em active particles}~\cite{example}.  These are generally
elongated and their direction of self-propulsion is set by their own
anisotropy (i.e., the two ends are distinguishable, hence {\em polar}),
instead of being determined by an externally imposed field. In contrast,
{\em active nematics}~\cite{sriram}, made of active particles which
are head-tail symmetric, do not show any self-propulsion. These active systems,
polar or nematic,   are generically characterised by the existence of
orientationally ordered states. These are nonequilibrium analogues
of the  equilibrium nematic liquid crystals.
There are numerous examples, which include both living systems ({\em living
matter}) as well as their artificially prepared non-living analogues.
Biological examples of active systems include both small and large in-vitro and
in-vivo systems, e.g., reconstituted bio-filaments and the associated
motor proteins~\cite{surrey}, the cytoskeleton of living cells and
bacterial suspensions~\cite{dombrowski}, cell
layers~\cite{kemkemer}, and also larger-size objects, e.g., flock of
birds or school of fishes~\cite{toner}. Analogous non-living examples of active
matter systems also arise in various contexts, e.g., layers of vibrated granular
rods~\cite{sriram-vijay} and colloidal or nanoscale particles propelled
through a fluid by catalytic activity at their
surface~\cite{paxton}. All these examples of active systems are distinguished
by a local energy supply in the bulk that drives the systems away from
equilibrium. This is in contrast to other well-known examples of driven systems,
e.g., sheared systems, where the external drives act at the boundaries. For
instance, in cell biology contexts, this supply of energy
takes place due to the hydrolysis of adenosine triphosphate (ATP) to
adenosine diphosphate (ADP) and other phosphates (Ph) by the molecular
motors, thus converting the chemical fuel into mechanical motion.

Despite hugely varying microscopic details, different active matter
systems display a host of intriguing nonequilibrium phenomena with
generic features independent of system details, e.g., pattern
formations, wave propagations, oscillations and unusually strong
fluctuations~\cite{sriram-aditi,sriram-aditi-toner,frey-nature}.
Due to the large number of
diverse microscopic variables present (especially in the cell biology
context), the level of complexities in active matter systems at microscopic levels is very
high. Instead, it is convenient to formulate the coarse-grained dynamics of the active systems
based on identifying global features, e.g., the presence or absence of conservation laws,
symmetries, the presence of appropriate broken symmetry variables
and the nature of the underlying momentum damping. These are similar
in spirit and nonequilibrium generalisation of the general principles
and laws developed to describe the statistical mechanics and
dynamics of the ordered phases in equilibrium systems~\cite{martin}.
These {\em active fluid} theories,
parametrised by a set of phenomenological
constants~\cite{aranson,kruse,sriram,vicsek,marchetti,joanny-rev},
serve as as generic coarse-grained
descriptions for a driven orientable fluid with nematic or polar symmetries
and are particularly useful to uncover and elucidate the long
wavelength behaviour observed in very different physical systems and at
very different length scales~\cite{toner,sriram-aditi,sriram-aditi-toner,pagla}.

 In a bulk fluid (both active and passive) the
viscosity damps out any local momentum gradient and thus reduces any
relative velocities between neighbouring regions. The
total momentum of the system is however kept conserved;
such systems are known as {\em wet active
matters} in the language of Ref.~\cite{marchetti}; see, e.g.
Refs.~\cite{sriram-aditi,rafael}. In contrast, for systems resting on a rigid
substrate (e.g., a layer of active fluid on a
solid substrate) there is a drag on the system acted typically
through a no-slip boundary condition on the active matter velocity
at the active matter-rigid substrate interface. This drag leads to
nonconservation of the  momentum of the active system and cuts off
any long-ranged hydrodynamic interactions. These are known as {\em dry active
matter} in the classification used in Ref.~\cite{marchetti} and have been
studied extensively, see, e.g.,
~\cite{toner,sriram-vijay,sriram-aditi-toner,shradha-sriram}. The properties of
active matter systems are often considered in the form of thin,
quasi two-dimensional (2D) layer. Such quasi-2D
active matter systems exist both {\em in-vivo} and {\em in-vitro}: cell
cortex~\cite{cortex} or the cortical actin layers and cell ruffles, e.g.,
lamellipodia~\cite{alberts} are examples belonging to the former category,
where as reconstituted actin layers on liposomes~\cite{joanny-cycil} are
examples of 2D {\em in-vitro} active fluid systems.

Inspired by the current studies on both wet and dry active matters
and their significant differences in terms of their long wavelength
properties, we study a generic 2D
 polar active matter layer, where the active particle system is
embedded inside a three-dimensional (3D) bulk isotropic passive
fluid. The active fluid and the embedding passive fluid interact via
a mutual friction at the interfaces of the active fluid-bulk fluid
interface, leading to momentum damping of the active particles. To
this end, we construct a set of 2D continuum equations of motion for
the local orientation and number density of the polar active
species. Our model, parametrised by the interfacial friction
$\Gamma$, provides a unified coarse-grained description of the
dynamics of polar ordered 2D wet and dry active matters and free
standing 2D films. In a linearised treatment about a chosen
orientationally ordered uniform state, we find the linear
instabilities in the system.  We also study the nematic limit of the
dynamics. The nature of the linear instabilities are found to depend
sensitively on the magnitude of $\Gamma$ relative to the viscous
damping.   Our results may be used to estimate bounds on $\Gamma$ in
possible physical realisations of our model, e.g.,  reconstituted
actin filaments deposited on a liposome embedded in a fluid medium.
In addition, in an {\em in-vivo} system of two eukaryotic cells with
a substantial area of contact, the dynamics of the cortical actin
layers of the two cells on both sides of the contact plane should be
describable by our dynamical equations at a coarse-grained level.
Nonetheless, our formulation is sufficiently general and does not
specifically relate to any particular cell biological example.  The
rest of the article is organised as follows: In Sec.~\ref{model}, we
define our model and set up the basic equations of motion. Then in
Secs.~\ref{highlin}, \ref{interlin} and \ref{weaklin}, we analyse
the instabilities for high, intermediate and low values of the
mutual friction. Then in Sec.~\ref{influgammaxx} we briefly compare
the linear instabilities in the different regimes of the model,
delineated by the magnitude of the mutual friction. In the next
Sec.~\ref{nemlim}, we analyse the nematic limit of our model
dynamics. We discuss and summarise in Sec.~\ref{summ}. Finally, we
provide some calculational details and then obtain the ambient
velocity profiles in the Appendices.

\section{Model Equations}
\label{model}

\begin{figure}[htb]
\includegraphics[height=8cm]{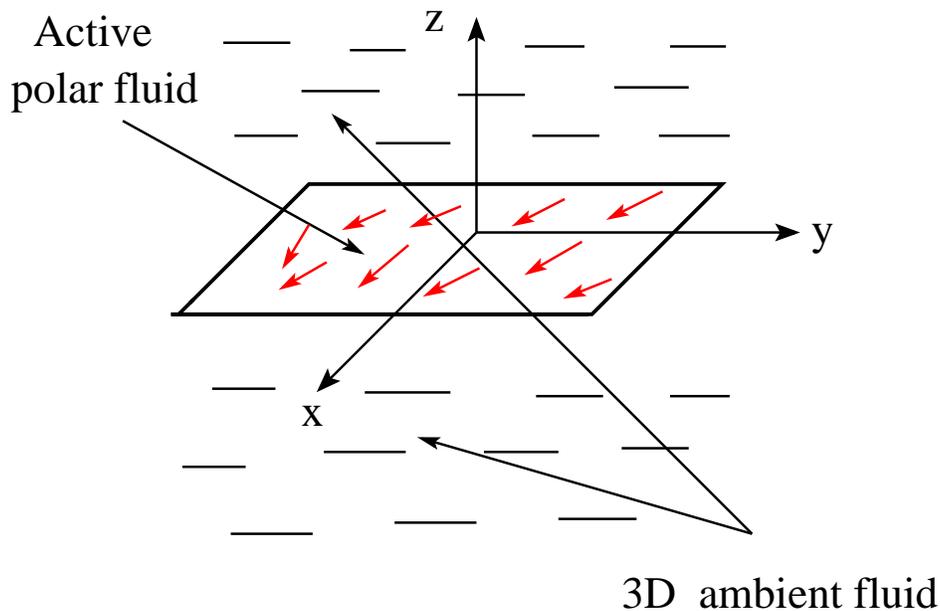}
\caption{(Color online) Schematic diagram of our model active fluid layer spread
along the $xy$-plane. Arrows indicate polar active particles, aligned
predominantly along the $x$-direction (see text).}
\label{modelfig}
\end{figure}

We consider an inflexible thin planar
layer of a viscous active fluid with a vanishingly small thickness,
located at the $xy$-plane, i.e., at
$z=0$. We treat it as a quasi 2D system, for which a 2D description
should be appropriate.  The local number densities of the active species and the
solvent are $\rho({\bf x})$ and $\phi({\bf x}),\,{\bf x}=(x,y)$,
respectively. The active fluid layer, with a 2D viscosity $\eta$, is embedded in a
3D passive incompressible ambient fluid with a 3D viscosity $\eta'$, both above ($z>0$) and
below ($z<0$); see Fig.~\ref{modelfig} for a schematic diagram of our
model system. It is not unusual to treat thin active fluid layers as
quasi-2D systems; see, e.g., Ref.~\cite{arnabsaha}. We expect this 2D
description with a 2D viscosity to be good
for really very thin system such that any variation of the physical quantities
along the thin direction may be neglected.

  The centre of mass velocity of
the  active particles and the solvent combined is given by ${\bf
v}$. The total number of both the active and solvent particles are
separately conserved: The continuity equations for $\rho$ and $\phi$
are written as \bea
\partial_t\rho + {\boldsymbol\nabla}\cdot {\bf J}_\rho &=& 0, \label{cont}\\
\partial_t\phi + {\boldsymbol\nabla}\cdot {\bf J}_\phi &=& 0.
\eea
 Here, ${\boldsymbol \nabla}\equiv \hat x\partial/\partial x +
 \hat y\partial/\partial y$ is the
 2D gradient
 operator, $\hat x,\,\hat y$ are the unit vectors along $x$- and
 $y$-directions.
  The particle currents ${\bf J}_\rho$ and ${\bf J}_\phi$ can be
expressed in terms of the 2D centre-of-mass velocity $\bf v$ and the
diffusion current $\bf j$. \bea
{\bf J_\rho} &=& \rho \bf v + j, \label{currrho}\\
{\bf J_\phi} &=& \phi \bf v - j. \label{currphi} \eea  Here, the
molecular masses of both the active and solvent
particles are assumed to be equal to unity for calculational convenience. We are
interested in an orientationally ordered state of the model system. To this end, we
introduce a 2D local polarisation vector ${\bf p}=(p_x,p_y)$, with a fixed
magnitude, $p^2=1$, as appropriate for an orientationally ordered
state. Microscopically, it describes the local orientations of the
actin filaments or bacteria. We consider the active fluid to be overall incompressible,
i.e., ${\boldsymbol\nabla}\cdot {\bf v}=0$. Our chosen reference
state is defined by $p_x=1$ with
no macroscopic overall flow, i.e., $\langle v_\alpha\rangle=0$. Note that
this does not rule out finite velocity of propagation (or a nonzero macroscopic current)
of the active particles (see below). In the
Stokesian limit of the flow dynamics, the force balance equation
\bea \nabla_\alpha\sigma_{\alpha\beta}
-\partial_\beta\Pi+F_\beta=0,\,{\rm with}\;\alpha,\beta=x,y \label{forcebal} \eea
 yields the generalised Stokes equation for $\bf v$.  The 2D pressure
$\Pi$ may be eliminated by using the incompressibility condition
($\nabla\cdot {\bf v}=0$). Here, $\sigma_{\alpha\beta}$ is the total
stress tensor and external forces $F_\beta$ are the tangential stresses of the
embedding fluid
 on the two sides (top and bottom) of the active fluid layer
\bea F_\beta=\eta'(\partial_z v'_\beta + \partial_\beta
v_z')|_{z=\epsilon} -\eta'(\partial_zv'_\beta + \partial_\beta
v_z')|_{z=-\epsilon}, \label{3dforce}
 \eea
  where $\epsilon\rightarrow
0$;  ${\bf v}'({\bf r})$ (with ${\bf r}=(x,y,z)$) is
the 3D  ambient fluid velocity.
    \begin{figure}[htb]
\includegraphics[height=8cm]{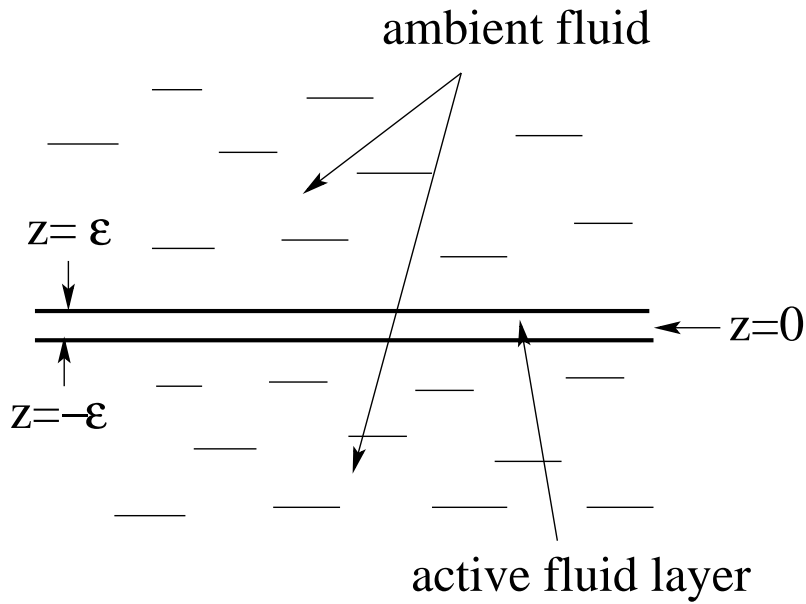}
\caption{(Color online) Schematic diagram of our model active fluid layer spread
along the $xy$-plane ($z=0$). The interfacial friction acts at
$z=\pm\epsilon,\epsilon\rightarrow 0$ (see text).}
\label{bc}
\end{figure}

In the spirit of linear response theories~\cite{martin}, the dynamics of the
active fluid layer is described in terms of linear relations between the
thermodynamic fluxes ($\sigma_{\alpha\beta}^s,\,j_\alpha,P_\alpha$) and
the corresponding generalised forces ($v_{\alpha\beta},\partial_\alpha
\overline\mu,h_\alpha$)~\cite{sriram-aditi,kruse,frank,frank2}.
 Here, $\sigma_{\alpha\beta}^s$ is the symmetric
part of the deviatoric stress
 \bea
\sigma_{\alpha\beta}^s=\sigma_{\alpha\beta}+\rho_t v_\alpha v_\beta
- \sigma_{\alpha\beta}^a , \label{totsigma}
\eea
 with $\sigma_{\alpha\beta}^a=(p_\alpha h_\beta -p_\beta
h_\alpha)/2$ is the antisymmetric part of the stress tensor, $h_\alpha$
being the thermodynamic force conjugate to polarisation $p_\alpha$. Further,
$\rho_t=\rho +\phi$ is the total density of the two species
combined and $v_{\alpha\beta}=(\partial_\alpha v_\beta +\partial_\beta v_\alpha)/2$
is local strain rate tensor. The term $\rho_t v_\alpha v_\beta$ is the Reynold's stress
in the active fluid.
In addition, ${\bf P}$ is the convected co-rotational derivative of the polarisation
vector given by \bea P_\alpha = {D \over Dt}p_\alpha = \partial_t
p_\alpha + v_\beta\partial_\beta p_\alpha
+\omega_{\alpha\beta}p_\beta, \label{orient} \eea  with
$\omega_{\alpha\beta}={1 \over 2} (\partial_\alpha v_\beta
-\partial_\beta v_\alpha)$ is the vorticity tensor. Furthermore,
$\bar\mu=\mu_\rho-\mu_\phi$ is the effective chemical potential;
$\mu_\rho$ and $\mu_\phi$ are individual chemical potentials of the
active particles and the solvent molecules, respectively. 
For simplicity we consider the dilute limit of the active
particles $\rho\ll \phi$ or $\rho_t\approx\phi$, i.e., ${\bf J}_\phi\approx \phi {\bf v}$.
In this limit, the overall incompressibility (which implies $\rho_t=const.$)
is equivalent to considering
$\phi=const.$, so that the dynamics of $\phi$ can be neglected and
we consider the dynamics of $\rho$ alone. In this dilute
limit, $\overline\mu$ may be replaced by the chemical potential $\mu_\rho$ for the active
particles.

The stress field is assumed to contain a nonequilibrium {\em active
stress} of the form \bea
\sigma_{\alpha\beta}^{act}=\zeta'(\rho)\Delta\mu p_\alpha p_\beta.
\label{actstress} \eea Microscopically, $\sigma_{\alpha\beta}^{act}$
is due to the local nonequilibrium dynamics of the active particles. The
coarse-grained form (\ref{actstress}) may be obtained~\cite{sriram-aditi} by
noting that the force applied by an active particle on the fluid surrounding it
is same as that applied by the fluid on it, considering the total forces
exerted by a collection of active particles with a given centre, each exerting
a point force proportional to and parallel to $\bf\pm p$ and expanding it up to
the lowest order in spatial gradients.
 Therefore, the magnitude of
$\sigma_{\alpha\beta}^{act}$ should depend on the local density
$\rho$ of the active particles; hence the form
$\zeta'=\zeta'(\rho)$. We write $\rho=\rho_0 +\delta\rho$, where
$\rho_0$ is the mean active particle density and $\delta\rho$ are
fluctuations (assumed small) about $\rho_0$. Expanding for small
$\delta\rho$, we write
$\zeta'(\rho)=\zeta+\overline\zeta\delta\rho$, where
$\zeta=\zeta'(\rho_0)$ and $\overline\zeta
=\partial\zeta'/\partial\rho|_{\rho=\rho_0}$. Parameter $\Delta\mu$
represents the strength of $\sigma_{\alpha\beta}^{act}$; the latter
 is said to be contractile or extensile depending on whether
$\Delta\mu$ is  negative or positive, respectively; $\Delta\mu$ is a
measure of the rate of supply of (free) energy that pushes the
system out of equilibrium; in the context of the cortical actins in a
cell, it is the hydrolysis of the ATP
 molecules to ADP and phosphates that supplies this energy;
 $\Delta\mu=\mu_{ATP}-\mu_{ADP}-\mu_{Ph}$ where
 $\mu_{ATP},\mu_{ADP},\mu_{Ph}$ are the chemical
 potentials of ATP, ADP and phosphate molecules.  Parameter $\Delta\mu$ has the
dimension of energy/(mass.mole). Numerical estimation of $\Delta\mu$ is
not easy: In the particular context of cell biology,
one may use the fact that approximately 7 kCal energy
released per mole of ATP due to its hydrolysis. Since 1
molar mass of ATP $\sim 500$, we obtain from its definition
$\Delta\mu\sim  7 kCal/(500gm/10^{23})$, the free energy release
per unit mass per molecule.


The relevant linear flux-force relations~\cite{frank,marchetti}, that include
the active stress contribution to the stress and allow for polar terms,
i.e., not invariant under $\bf p\rightarrow -p$, are
 \bea \sigma_{\alpha\beta}^s &=&
2\eta v_{\alpha\beta} + \zeta'(\rho)\Delta\mu p_\alpha p_\beta
+ {\nu_1 \over 2}(p_\alpha h_\beta + p_\beta h_\alpha)-{\epsilon_0
\over 2} (p_\alpha\partial_\beta\bar\mu_\rho +p_\beta\partial_\alpha\bar\mu_\rho),
\label{sigmatless}\\
j_{\alpha} &=& -\gamma_{\rho\rho}\partial_\alpha\mu_\rho
+\bar\lambda h_\alpha + \kappa_\rho p_\alpha\Delta\mu +w\Delta\mu
\partial_\beta(\rho p_\alpha p_\beta) - {\epsilon_0 \over
2}p_\beta(\partial_\alpha v_\beta +\partial_\beta v_\alpha),
\label{jrho} \\
P_\alpha &=& {h_\alpha \over \gamma_0} + \lambda_1p_\alpha \Delta\mu
+ \nu_1 p_\beta { v}_{\alpha\beta} -\overline\lambda\partial_\alpha\overline\mu_\rho + \lambda_2({\bf p}\cdot
\nabla)p_\alpha\Delta\mu + \lambda_\rho
\partial_\alpha \rho \Delta\mu. \label{pcoder}
\eea Coupling constant $\nu_1$ denotes the equilibrium
flow-orientation coupling~\cite{jacques-book}; similarly,
$\epsilon_0$ denotes symmetry-allowed equilibrium couplings between
the flow and the particle current~\cite{frank}.  Parameter
$\gamma_{\rho\rho}>0$ is a mobility coefficient (an equilibrium
coupling constant) and related to the diffusion coefficient.  In
addition, particle current ${\bf j}_\rho$ should have active
contributions   $\kappa_\rho\Delta\mu {\bf p}$ and $w\Delta\mu
\partial_\beta(\rho p_\alpha p_\beta)$, such that there should be an
active macroscopic current of the particles in the direction
of $\bf p$, with  amplitudes proportional to $\Delta\mu$. In
addition, $\overline\lambda$ is a cross-coupling equilibrium
coupling constant. (In general, $\overline\lambda$ may be a tensor
reflecting the anisotropy of the polar ordered state; we neglect
this here.) Notice that in (\ref{pcoder}) we include two
symmetry-permitted active terms with coefficients $\lambda_2$ and
$\lambda_\rho$, respectively; the $\lambda_2$-term is a
self-advection term, (not considered in Refs.~\cite{frank,frank2}). 
Since the active particles tend to display macroscopic motion
with respect to the embedding fluid even in their fully ordered state (no
distortion), microscopically the $\lambda_2$-term represents advection of the
local distortions in $\bf p$ by $\bf p$. The
$\lambda_\rho$-term is a nonequilibrium
partial pressure term, modelling motion of the active particles along
or opposite to the concentration gradient (depending upon the sign).
Coefficients $\zeta',\kappa_\rho,\lambda_1,\lambda_2, w$ and
$\lambda_\rho$ are "active coefficients", i.e., coefficients of
different active terms in Eqs.~(\ref{sigmatless}-\ref{pcoder}). Out
of these, $\kappa_\rho,\lambda_2$ and $\lambda_\rho$ are
coefficients of the different polar terms, which break the symmetry
under $\bf p\rightarrow-p$, where as $w,\zeta$ and $\overline\zeta$
are coefficients of the nematic active terms in the dynamics. Thus,
in the nematic limit of the model, $\kappa_\rho,\lambda_2$ and
$\lambda_\rho$ are all zero, and the only source of nonequilibrium
drive is the active stress (\ref{actstress}) and the active particle
current represented by the $w$-term.  For reasons similar to the
$\rho$-dependence of $\zeta'$, active coefficient $\kappa_\rho$
should depend on $\rho$. We write for small density
fluctuations~\cite{sriram-aditi}
 \bea
\kappa_\rho(\rho) =\kappa_0
+\kappa_{\rho\rho}\delta\rho,\;\kappa_{\rho\rho}
=\frac{\partial\kappa_\rho}{\partial\rho}|_{\rho=\rho_0},
\eea where $\kappa_0=\kappa_\rho (\rho_0)$ depends on the mean
density and $\kappa_{\rho\rho}$ incorporates the effects of the
fluctuations of $\rho$ about $\rho_0$. We ignore any
$\rho$-dependence of $\lambda_\rho$ and $\lambda_2$ and treat them
as a constant, since we are interested in a linearised treatment.

 Thermodynamic forces
$\bf h$ and $\mu_\rho$ are defined as follows \bea
h_\alpha=-\frac{\delta \mathcal F_0}{\delta
p_\alpha},\;\mu_\rho=\frac{\delta \mathcal F_0}{\delta \rho},
\label{thermoforce}
\eea
where $\mathcal F_0$ is a free energy functional that controls the
relaxation of the system to its thermal equilibrium state in the absence of any activity.
 At the bilinear
order in fields \bea \mathcal{F}_0=\int d^2x\frac{1}{2}[D({\nabla_\alpha}
p_\beta)^2 + A(\delta \rho)^2 + 2\chi\rho{\boldsymbol\nabla}\cdot {\bf p}],
\label{free1} \eea
where $D$ is a 2D Frank elastic constant (we have assumed
equal Frank's constants for simplicity),
$A\sim T\rho_0$ is an osmotic modulus with $T$ being the temperature
when the system is in thermal equilibrium, $\chi$ provides a
symmetry-allowed coupling between the
density fluctuations and splay. Assuming the minimum free energy
configuration to be given by a uniform configuration ${\bf
p}=const.$ and $\rho=\rho_0$ everywhere, we must have $\chi^2 < A D$.
From (\ref{free1}), we find
$h_\alpha=-{\delta\mathcal{F}_0 \over \delta p_\alpha} =
D\nabla^2p_\alpha +\chi\partial_\alpha\rho$ and
$\mu_\rho=\frac{\delta\mathcal{F}_0}{\delta (\delta\rho)}=
A\delta\rho+\chi {\boldsymbol\nabla}\cdot {\bf p}$.
Eliminating $\Pi$ and using the forms of $\bf h$ and
$\mu_\rho$, we obtain to the lowest order in spatial
gradients (see Appendix~\ref{stokesderi})
 \bea \eta\nabla^2v_\beta + \zeta\Delta\mu
P_{\beta\gamma}\partial_x p_\gamma + \zeta \Delta\mu P_{\beta
x}\partial_yp_y +\bar\zeta\Delta\mu P_{\beta x}\partial_x\rho =
-F_\beta, \label{stokes} \eea where we have linearised about
$p_x=1$, $\rho=\rho_0$.

For an isotropic, passive ambient fluid medium, in the low Reynolds
number limit and for small masses, its velocity $v'_i,\,i=x,y,z$
satisfies the Stokes Eq.:
 \bea \eta'\nabla_3^2 v'_i = \nabla_{3i}\Pi',
\label{stokessubsup}
 \eea
  valid for both the super- ($z>0$) and sub-
($z<0$) phases, $\Pi'$ is the ambient fluid pressure and
${\boldsymbol\nabla}_3$ is the 3D gradient operator. The boundary
conditions on $v_i'$ are as follows:
\begin{itemize}
\item No flow at infinity: At both
$z\rightarrow\pm\infty$, $v'_i$ should vanish.
\item  Balance of the normal
stresses of the ambient fluid at the 2D active fluid layer,
  \item Due to the assumed inflexibility of the active fluid layer,
  the normal velocity of the ambient fluid at the
active fluid layer should be zero:
$v_z'(z=\epsilon)=0=v_z'(z=-\epsilon)$.
\item Boundary conditions at the active fluid-bulk fluid interfaces
requires careful consideration; see Fig~\ref{bc}.
The most common boundary condition used in this context is
the "no-slip" condition, i.e., equality of the active fluid
velocity and the in-plane component of the 3D ambient fluid velocity at  the top and bottom
interfaces between the ambient fluid and the active fluid layer.
We generalise this by allowing a slip. We implement this by introducing
a {\em slip coefficient of friction}, such that the
shear stresses are balanced by the friction forces at the
interfaces. This implies (using $v_z'=0$ at $z=\pm\epsilon$)
 \bea \eta'\partial_zv'_\beta|_{z=\epsilon} &=& \Gamma
(v'_\beta |_\epsilon - v_\beta), \label{fricup1} \\
\eta'\partial_zv'_\beta|_{z=-\epsilon} &=&
-\Gamma (v'_\beta |_{-\epsilon} - v_\beta), \,\beta=x,y\label{fricdown1} \eea
where $\Gamma$ is the slip coefficient of friction at the upper and lower
interfaces (we assume equal friction  at the upper and
lower interfaces for simplicity);
this allows us to define a {\em slip length} $l_s\sim \eta'/\Gamma$. 
 \end{itemize}
 Notice that for a finite $\Gamma$,
boundary conditions (\ref{3dforce1}) implies partial slip between
$v_\beta (x,y)$ and $v'_\beta(x,y,z=\pm\epsilon),\,\beta=x,y$. 
While the no-slip boundary condition is more
conventionally used, on mesoscopic scales, however, instances of
violation of the no-slip boundary conditions are known. For instance,
Ref.~\cite{zhu} has shown that beyond a critical shear stress that
depends strongly on the surface roughness, departure from the no-slip
conditions may be observed. It has also been found that upon
addition of surfactant in the fluid, the boundary condition changes from
no-slip to partial slip~\cite{zhu1}. In addition, there are now strong
evidences in favour of slip in polymer melts; see, e.g. Refs.~\cite{zhao,lam}. Furthermore,
it has been demonstrated in Ref.~\cite{ehlinger} how a large slip at a
liquid-liquid interface may be introduced experimentally. Friction has
been considered in various active fluid flow problems as well; see, e.g.,
Refs.~\cite{joanny1,joanny2,joanny3,frank1,yeomans} for various
theoretical and expreimental studies.
While no systematic measurements of slip at interfaces involving
active fluids are known, the above existing results suggest that
considering the complex internal structure of the active fluid
(e.g., the presence of actin filaments), a partial slip at the
interfaces between the active fluid layer and the 3D embedding fluid
cannot be ruled out. Recently, it has been shown that
a significant reduction of the sliding frictional forces between two bundled
F-actine filaments may be achieved by coating the F-actins with
polymeric brushes~\cite{dogic}.
Thus it is important to study implications of
finite slips in an active fluid problem in a simple set up, which we set out to do
below by using our model system. Notice that vanishing
$v_z'$ at $z=\pm\epsilon$ implies that the
shear forces $F_\beta$ on the active fluid layer as given in
(\ref{3dforce}) take the simpler form \beq
F_\beta=\eta'(\partial_zv_\beta'|_{z=\epsilon} -
\partial_zv_\beta'|_{z=-\epsilon}). \label{3dforce1} \eeq
 By using the boundary conditions prescribed
above, together with the incompressibility of the ambient fluid
${\boldsymbol \nabla}_3\cdot {\bf v}'=0$, Stokes'
Eq.~(\ref{stokessubsup}) may be solved to yield $v_i',\,i=x,y,z$ (see
Appendix) and obtain $F_\beta$.

Equation of motion for the orientational field $p_\alpha$ may be
written combining equations (\ref{orient}) and (\ref{pcoder}). \bea
\partial_t p_\alpha + v_\beta\partial_\beta p_\alpha
+\omega_{\alpha\beta}p_\beta = {h_\alpha \over \gamma_0} +
\lambda_1p_\alpha \Delta\mu + \nu_1 p_\beta { v}_{\alpha\beta} -
\bar\lambda \partial_\alpha \mu_\rho + \lambda_2({\bf p}\cdot
{\boldsymbol\nabla})p_\alpha\Delta\mu + \lambda_\rho
\partial_\alpha \rho \Delta\mu. \label{orientfull}
\eea With $p_x=1$ defining the reference state, $p_y$ is a broken
symmetry (slow) mode. We linearise (\ref{orientfull}) above about
$p_x=1$ for small $p_y$. This yields
 \bea {\partial_t p_y} =
{(D\nabla^2p_y+\chi\partial_y\rho) \over \gamma_0} +
\lambda_2\Delta\mu\partial_xp_y +
\lambda_\rho\Delta\mu\partial_y\rho + {(\nu_1-1) \over
2}\partial_yv_x +{(\nu_1+1) \over 2}\partial_xv_y-\bar\lambda A
\partial_y\rho-\bar\lambda\chi\partial_y^2p_y. \label{py}
 \eea
Note that in Eq.~(\ref{py}), $\rho$ enters into the dynamics of
$p_y$ through  both equilibrium and
nonequilibrium contributions. Both are equally relevant being
the lowest order terms in gradient expansions.

The equation of motion for $\rho$ is obtained by using
Eqs.~(\ref{cont}), (\ref{currrho}) and (\ref{jrho}). Up to the order
$q^2$ the equation of motion for $\rho$ in the Fourier space,
linearised about $p_x=1$, is obtained as (set $A=1$)
 \bea
\partial_t\rho = -\gamma_{\rho\rho}q^2\rho +w\Delta\mu q_x^2\rho
+w\Delta\mu\rho_0q_xq_y\rho
-i\Delta\mu\kappa_0q_yp_y -i\Delta\mu\kappa_{\rho\rho}q_x\rho
+iD\bar\lambda q_y q^2p_y+\bar\lambda\chi q_y^2 \rho,
\label{rhoeq} \eea
 where ${\bf q}=(q_x,q_y)$ is the in-plane Fourier wavevector, conjugate to
 ${\bf x}=(x,y)$.

 Notice that in the linear equations (\ref{stokes}), (\ref{py}) and
 (\ref{rhoeq}) there are seven active coefficients (excluding
 $\Delta\mu$), which are introduced in the standard active fluid
models~\cite{joanny-rev,marchetti,kruse}.
 Out of these,  $\lambda_2,\lambda_\rho,
\kappa_0$ and $\kappa_{\rho\rho}$ control the  conditions for
instabilities (along with the sign of $\Delta\mu$) for both the high
friction and intermediate friction cases (see below).
 In terms of an underlying equivalent agent-based microscopic dynamics, we expect all
 these coefficients to depend upon the local density of the active
 particles and the specific alignment rules (favouring nematic or polar
 alignment). Thus, it is reasonable to expect that all the seven active coefficients
 are not independent parameters. On dimensional ground we argue that the two
 nematic active coefficients $\zeta$ and $\overline\zeta$ should be
 related as $\overline\zeta\sim \rho_0\zeta$ and the pairs of polar
 active coefficients in the active particle current $(\kappa_0,\kappa_{\rho\rho})$
 and in the active alignment
 $(\lambda_2,\lambda_\rho)$ are related as
 $\kappa_{\rho\rho}\rho_0\sim\kappa_\rho$ and
 $\lambda_\rho\rho_0\sim\lambda_2$, respectively. With the expectation that the
 polar alignment polar and active current terms originate from same
 underlying (polar) microscopic rules, we expect them to be mutually
 simply related. Again on dimensional ground we expect
 $\lambda_2\sim \kappa_{\rho\rho}$. Note however that in the all the
 above heuristic relations, there are dimensionless proportionality
 constants which we cannot obtain on simple physical
 ground. We would like to emphasise that all these parameters are just phenomenological
 constants, similar to the parameters
which appear in the continuum theories to describe the statistical
mechanics and dynamics of the ordered phases in equilibrium
systems~\cite{martin} and cannot be calculated within our theory. It
should in-principle be possible to relate these coefficients to and
calculate them from the specific microscopic rules for agent based
models for active systems; see, e.g.,
Refs.~\cite{chate1,bertin2,erwin}. These are, however, outside the
scope of the present study. In what follows below, we ignore this
issue for simplicity and treat all the seven coefficients as
 independent model parameters. In addition to these seven active coefficients,
 the friction coefficient $\Gamma$ is not an active coefficient. This
enters into the dynamics through the boundary conditions and is a
model
parameter that has been introduced by us and is central to the
present discussion.
To our knowledge, no good estimate about the magnitude of $\Gamma$ is available;
we therefore treat $\Gamma$ as a free parameter in our model. In general,
Eqs.~(\ref{stokes}), (\ref{py}) and (\ref{rhoeq}) can be solved in principle
for arbitrary
values of $\Gamma$. Nonetheless, it is instructive to consider three different
limits of $\Gamma$, characterised by $l_s,\eta'$, system size $L$ and thickness
$d$ of the 2D active system, and analyse them separately, as discussed below.

\subsection{High friction limit}

We consider a "large" $\Gamma$: $\Gamma \gg\eta'/L$. Formally, we consider the
dynamics in the limit
$\Gamma\rightarrow\infty$ (equivalently, $l_s\rightarrow 0$); this is valid for
wavectors $ql_s\ll 1$, or for a system of linear size $L$, $l_s/L\ll 1$. Thus,
the system size must be much larger than the slip length. The stress balance
equations
(\ref{fricup1}) and (\ref{fricdown1}) yield
 \bea v_\alpha'|_{z=\epsilon} =
v_\alpha'|_{z=-\epsilon}=v_\alpha,\;\alpha=x,y. \label{velcont} \eea
Thus, there is no slip between the ambient fluid velocity at the
active fluid layer $v_\alpha'|_{z=\pm\epsilon}$ and the active fluid
velocity $v_\alpha$. Equation~(\ref{velcont}) forms one of the
boundary conditions on the ambient fluid velocity $v'_\alpha$.

Forces (shear stresses) $F_\alpha$ then may be expressed as (see
Appendix~\ref{fbetahigh};
{\bf see also Ref.~\cite{shear}}), \bea
F_x &=& -2qv_x\eta', \label{fx}\\
F_y &=& -2qv_y\eta', \label{fy}
\eea
where ${\bf q}=(q_x,q_y)$ is the in-plane Fourier wavevector. Putting
the values of (\ref{fx}) and (\ref{fy}) in the Stokes
equation (\ref{stokes}), the expressions for $v_x$ and $v_y$ can be derived
up to the lowest order in ${\bf q}$ linearising about $p_x=1$ and $\rho=\rho_0$.
\bea
v_x &=& -i{\zeta q_x^2q_y \over 2\eta'q^3}\Delta\mu p_y + i
{\zeta q_y^3 \over 2\eta'q^3}\Delta\mu p_y + i{\bar\zeta
q_y^2q_x \over 2\eta'q^3}\Delta\mu\rho, \label{vxfinal} \\
v_y &=&  -i{\zeta q_y^2q_x \over 2\eta'q^3}\Delta\mu p_y + i {\zeta
q_x^3 \over 2\eta'q^3}\Delta\mu p_y - i{\bar\zeta q_x^2q_y \over
2\eta'q^3}\Delta\mu\rho. \label{vyfinal} \eea
 Thus, $v_\alpha$ at $O(q^0)$ has only active contributions.

Equation (\ref{py}) may be written by substituting  for $v_x$ and
$v_y$ from Eqs.~(\ref{vxfinal}) and (\ref{vyfinal}). We thus
obtain
 \bea
\partial_tp_y &=& \frac{-Dq^2p_y + i\chi q_y\rho}{\gamma_0} +
i\lambda_\rho\Delta\mu q_y\rho +i\lambda_2\Delta\mu
q_xp_y-i\bar\lambda q_y\rho-\bar\lambda\chi q_y^2p_y\nonumber \\&&
 -{1 \over 4\eta'q}[(\nu_1-1)q_y^2-(\nu_1+1)q_x^2] \left[{\zeta
\Delta\mu  }\left(1- {2q_x^2 \over q^2}\right)p_y +
{\bar\zeta\Delta\mu q_xq_y \over  q^2}\rho\right], \label{pyeq} \eea
in the Fourier space.

\subsection{Intermediate friction}

For intermediate values of $\Gamma$, there are a
considerable slip between the ambient fluid velocity
$v_\alpha'|_{z=\pm\epsilon}$ and the active fluid velocity
$v_\alpha,\,\alpha=x,y$.
We consider the limit $q\gg \Gamma/\eta'$ (equivalently $q\gg l_s^{-1}$),
(\ref{fricup1}) and (\ref{fricdown1}) reduce to (see Appendix~\ref{fbetainter})
\bea
\eta'\partial_zv'_\beta|_{\epsilon} &=& -\Gamma v_\beta, \label{fricweakup} \\
\eta'\partial_zv'_\beta|_{-\epsilon} &=& \Gamma v_\beta, \label{fricweakdown}.
\eea
Clearly, these would be valid for wavevector $ql_s\gg 1$ or a system with size
$L \ll l_s$. In addition, we should have $\eta q^2 \ll \Gamma$. Since
$\eta\sim \eta' d$, where $d$ is the thickness of the 2D active fluid layer, we
obtain $L\gg (l_s d)^{1/2}$, yielding $l_s\gg L\gg (l_s d)^{1/2}$.

Substituting  (\ref{fricweakup}) and (\ref{fricweakdown}) in  (\ref{stokes})
and linearising about $p_x=1$ and $\rho=\rho_0$, the generalised Stokes
equations for $v_x$ and $v_y$ are obtained as (see Appendix~\ref{fbetainter})
\bea
v_x &=& i{\zeta \Delta\mu q_y \over 2\Gamma}\left(1- {2q_x^2
\over q^2}\right)p_y + i{\bar\zeta\Delta\mu q_xq_y^2 \over 2\Gamma q^2}\rho,
\label{vxweak} \\
v_y &=& i{\zeta\Delta\mu q_x \over 2\Gamma}\left(1- {2q_y^2 \over
q^2}\right)p_y - i{\bar\zeta\Delta\mu q_x^2q_y \over 2\Gamma
q^2}\rho, \label{vyweak} \eea
 where we have neglected $\eta q^2
v_\alpha$ in the limit $\eta q^2 \ll 2\Gamma$.
 This should be valid in the wavevector range satisfying
$\eta q^2 \ll \Gamma \ll\eta'q$. With $\eta\sim \eta'd$, taking $d\sim
10^{-7}m$ for a cortical actin layer and $\eta'\sim 10^{-3} N.sec/m^2$ for water, the above
inequality should hold over a wide range of $q$. As before, $v_\alpha$
has only active contributions at the lowest order in $q$.
Similar to an ordered active polar fluid layer on a solid substrate,
the
hydrodynamic interactions here are completely cut off  by the friction
$\Gamma$ and consequently ${\bf v}\sim O(q)$ to the lowest order in the
wavevector. Not surprisingly, Eqs.
(\ref{vxweak}) and (\ref{vyweak}) are identical in structure with the form
of the velocities of an active polar fluid layer resting on a solid surface.
This is due to the fact that for $ql_s\gg 1$, $v_i',\,i=x,y$ are effectively
very small and hence ignored. This background fluid thus effectively
behaves as a fixed background with the force on the 2D flow being given by
$-\Gamma v_\alpha$ (similar to a rigid substrate). Thus, with an intermediate
value for $\Gamma$, our model active system  corresponds surprisingly to a {\em
dry active matter}, despite being in contact with an embedding bulk fluid.

Using the above Eqs.~(\ref{vxweak}) and (\ref{vyweak}) in
eq.~(\ref{py}), the equation for $p_y$ can be written as \bea
\partial_tp_y &=& i(\frac{\chi}{\gamma_0}+\lambda_\rho \Delta\mu)q_y\rho
-{1 \over 4\Gamma}[(\nu_1-1)q_y^2-(\nu_1+1)q_x^2]
\left[{\zeta \Delta\mu  }\left(1- {2q_x^2
\over q^2}\right)p_y + {\bar\zeta\Delta\mu q_xq_y \over
q^2}\rho\right]
\nonumber \\
&&-{Dq^2p_y \over \gamma_0}
+ i\lambda_2\Delta\mu q_xp_y-i\bar\lambda q_y\rho-\bar\lambda\chi q_y^2p_y. \label{pyweak}
\eea


Density $\rho$ of the active particles still follows
Eq.~(\ref{rhoeq}). Notice that
Eqs.~(\ref{pyweak}) and (\ref{rhoeq})  are the linearised version of the model Eqs.
for a polar flock in Ref.~\cite{toner}, which is a coarse-grained
model for an active polar flock in a frictional medium. Thus, with
$(l_sd)^{1/2}\ll L\ll l_s$,
the long wavelength dynamics of our model is identical to that of a
polar-ordered layer of a suspension of active particles on a solid substrate, an
example of {\em dry active matters}. In this regime, our model is a
representation of Ref.~\cite{toner}.

\subsection{Weak friction limit}

In this case, $\Gamma$ is so small that $\eta q^2 \gg\Gamma$, or,
$q^2\gg \Gamma/\eta\sim\Gamma/(\eta' d)\sim 1/(l_s d)$; equivalently, $L\ll (l_s
d)^{1/2}$. Since $d$ is small for a quasi-2D system, $l_s$ must be very large
or $\Gamma$ very small for a physical system to display the weak friction
limit. From
the generalised Stokes Eq. for $\bf v$ (valid now for system size $L< (\eta/\Gamma)^{1/2}$)
we find
 \bea v_x &=& -i{\zeta
q_x^2q_y \over \eta q^4}\Delta\mu p_y + i {\zeta q_y^3 \over \eta
q^4}\Delta\mu p_y + i{\bar\zeta
q_y^2q_x \over \eta q^4}\Delta\mu\rho, \label{vxfinalweak} \\
v_y &=& - i{\zeta q_y^2q_x \over \eta q^4}\Delta\mu p_y + i {\zeta
q_x^3 \over \eta q^4}\Delta\mu p_y - i{\bar\zeta q_x^2q_y \over \eta
q^4}\Delta\mu\rho. \label{vyfinalweak} \eea
 Thus, ${\bf v}\sim O(1/q)$ at the lowest order, in contrast to
 the $q$-dependences of the velocities  for large or intermediate
 $\Gamma$ above. The differences are due to the lack of any screening
 of the hydrodynamic interactions in the present case.
Effectively, in this limit, the active fluid layer is a free
standing system being completely decoupled dynamically from the
ambient fluid. The dynamical equation for $p_y$ takes the form
 \bea
\partial_t p_y &=& i(\frac{\chi}{\gamma_0}+\lambda_\rho \Delta\mu)q_y\rho
- \frac{1}{2\eta q^2}[(\nu_1-1)q_y^2-(\nu_1+1)q_x^2]\left[{\zeta
\Delta\mu }\left(1- {2q_x^2 \over q^2}\right)p_y +
{\bar\zeta\Delta\mu q_xq_y \over  q^2}\rho\right]
\nonumber \\
&&-{Dq^2p_y \over \gamma_0} + i\lambda_2\Delta\mu q_xp_y-i\bar\lambda q_y\rho-\bar\lambda\chi q_y^2p_y.
\label{psiweak1}
 \eea
  see, e.g., Ref.~\cite{sriram-aditi}. Equation of
motion of $\rho$ is still given by Eq.~(\ref{rhoeq}).

\section{Linear Instabilities}

We now
 analyse the linear stability of the system from the dynamical
 equations obtained above by assuming a time-dependence for $p_y$
 and $\rho$ of the general form $\exp(\Lambda t)$.
  There are two independent modes, which may be static or moving, stable
  or unstable, given by  two values of $\Lambda$.  We calculate $\Lambda$
up to the lowest order in wavevector ${\bf q}$ for the different cases elucidated above.

\subsection{High friction limit}\label{highlin}



\begin{figure}[htb]
\includegraphics[height=6cm]{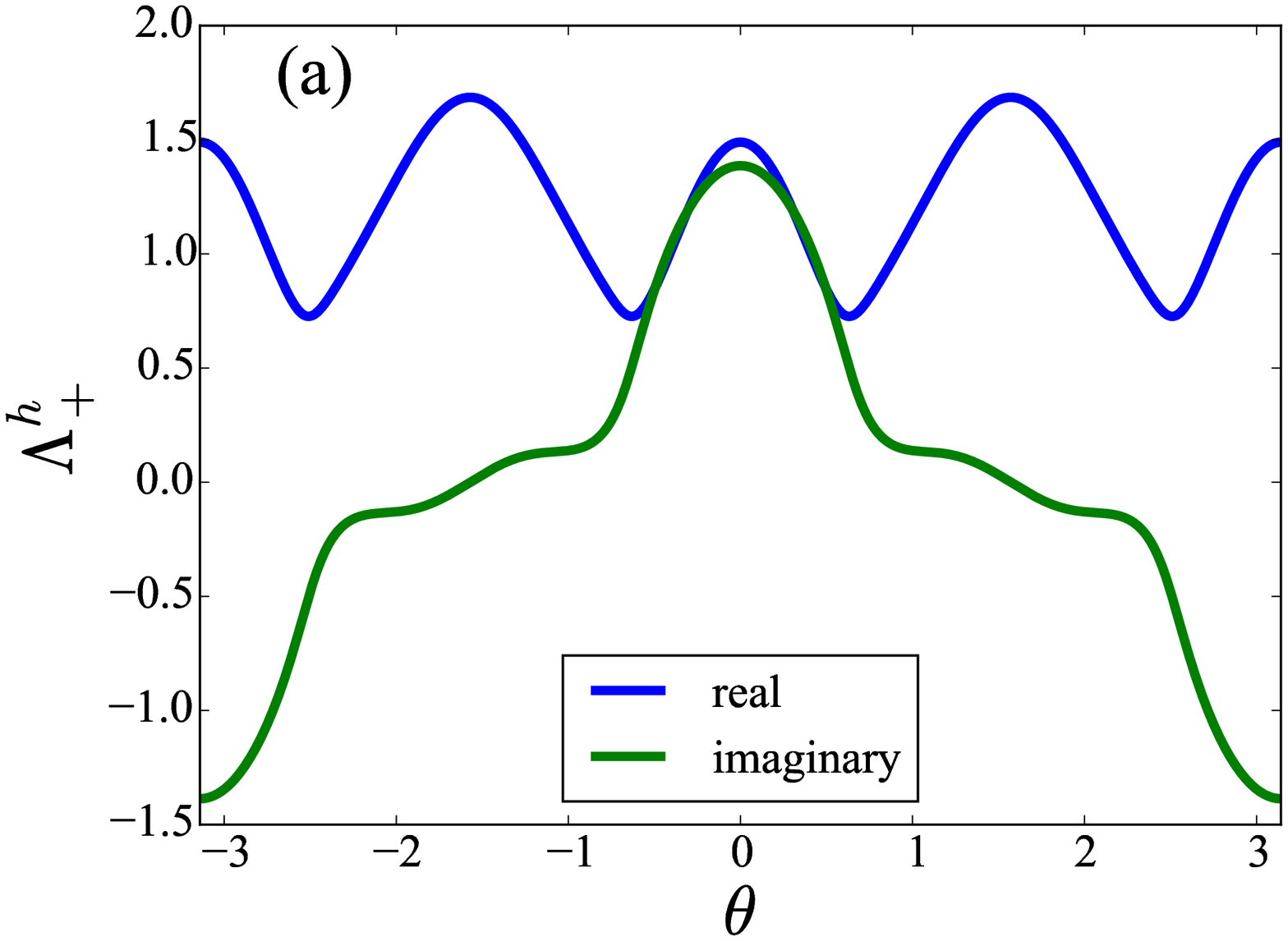},
\includegraphics[height=6cm]{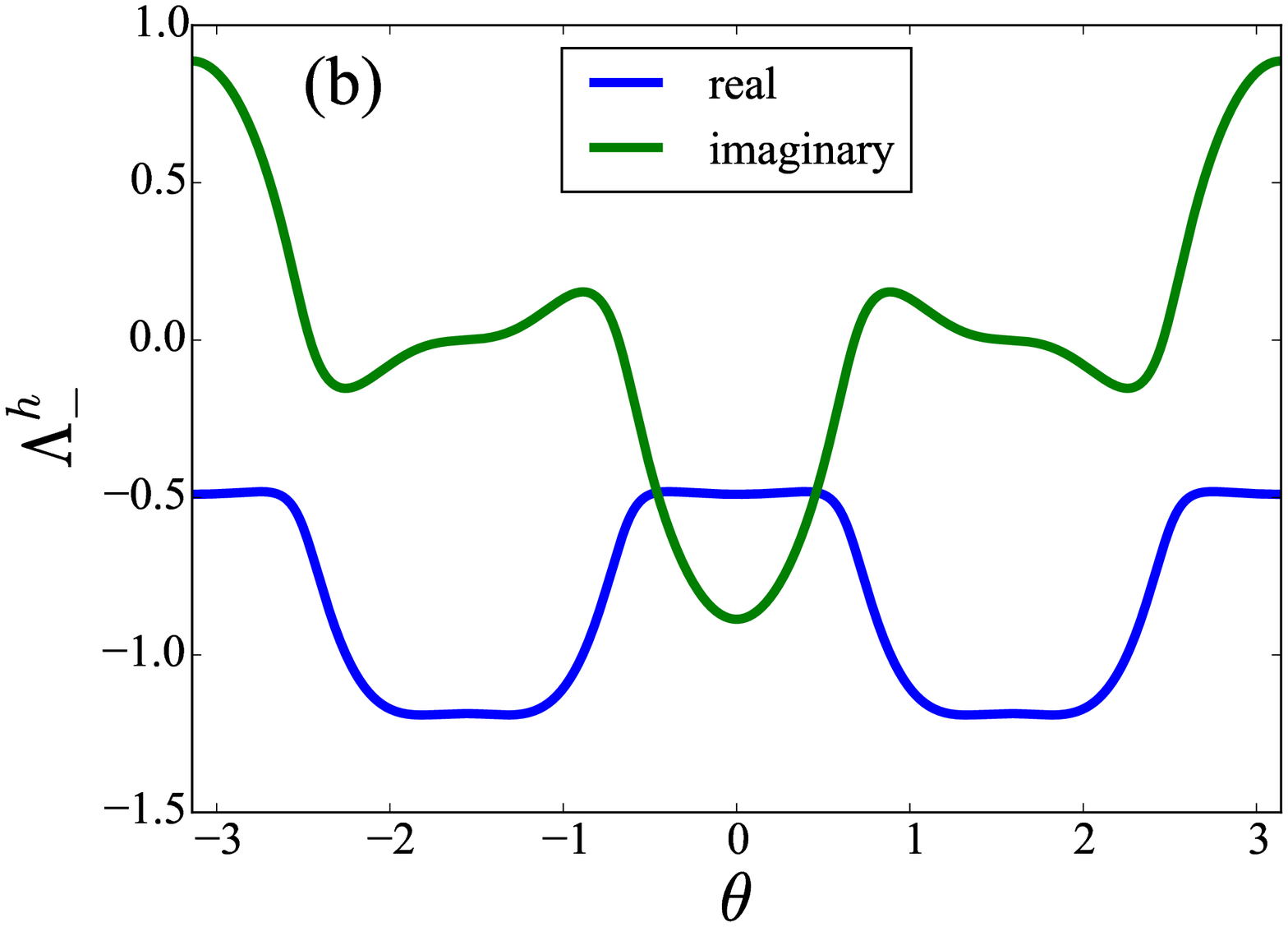}\\
\includegraphics[height=6cm]{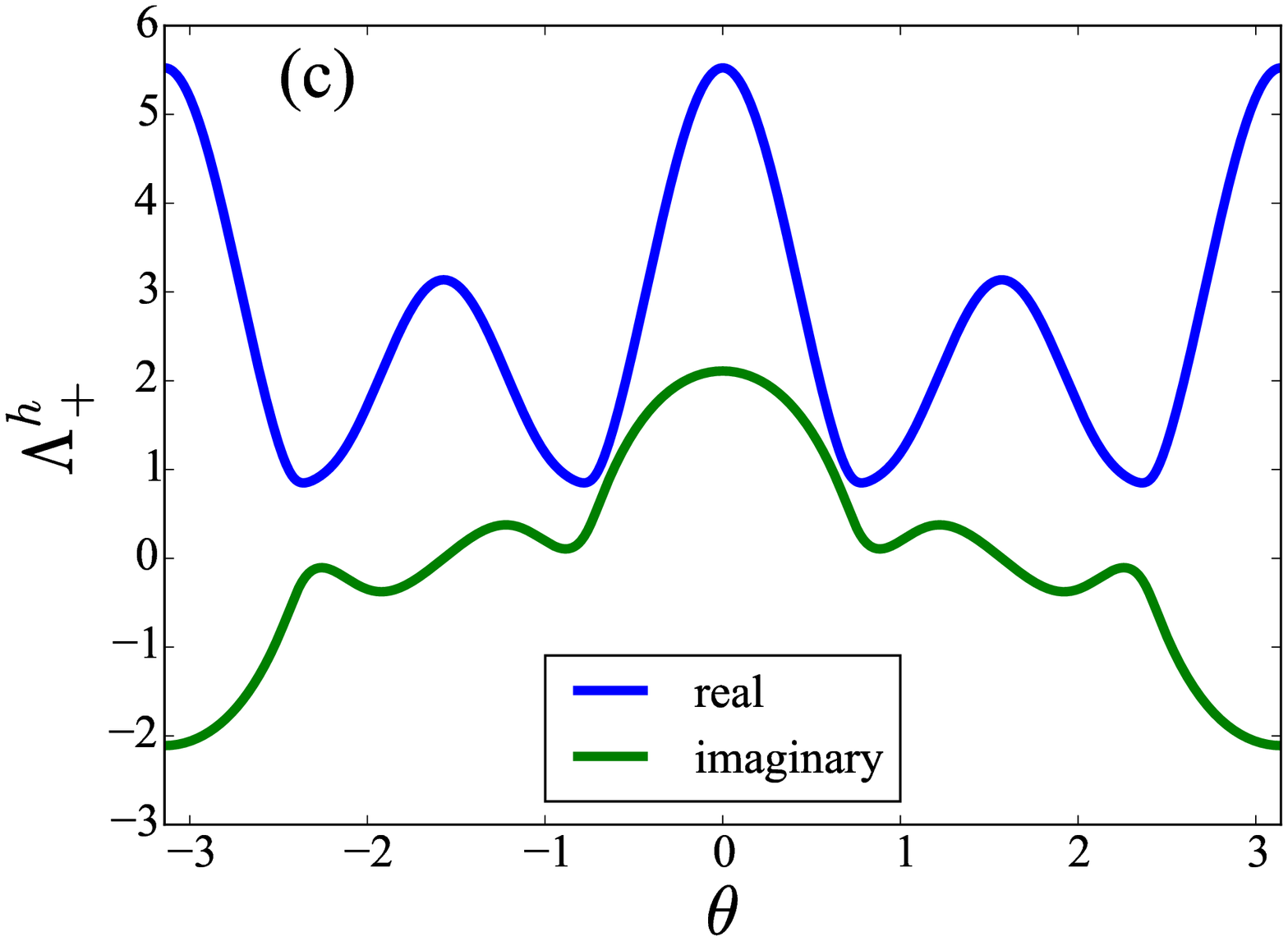},
\includegraphics[height=6cm]{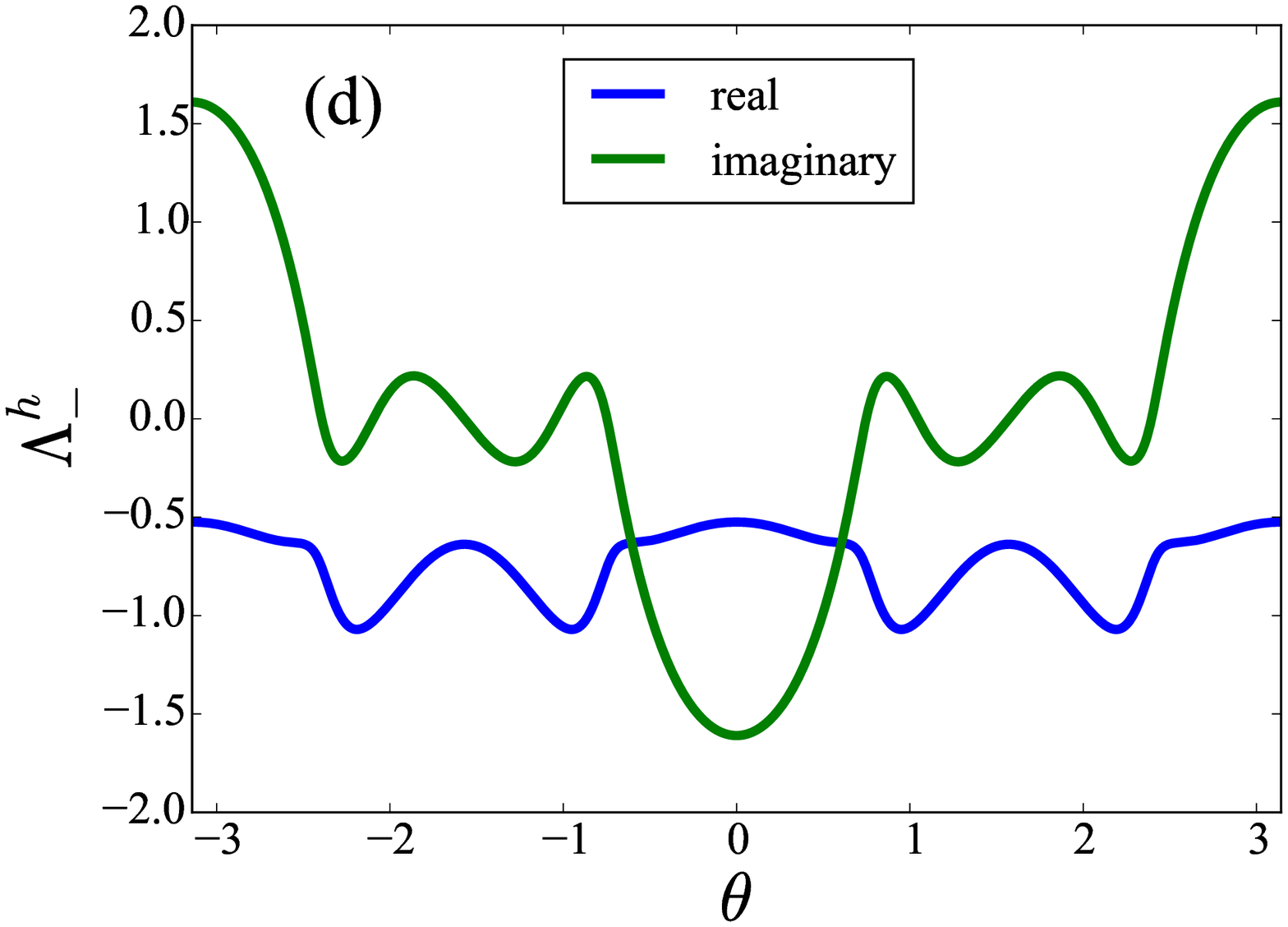}\\
\caption{(Color online) Representative plots of the real (blue line) and imaginary
(green line) parts of (a) eigenmode $\Lambda^h_+$ vs $\theta$ for $\zeta=1$,
(b) eigenmode $\Lambda^h_-$ vs $\theta$ for $\zeta=1$,
(c) eigenmode $\Lambda^h_+$ vs $\theta$ for $\zeta=5$, and
(d) eigenmode $\Lambda^h_-$ vs $\theta$ for $\zeta=5$
with fixed values of the other parameters; $\lambda_2=1, \kappa_{\rho\rho}=1/2,
\Delta\mu=1, \nu_1=3, \eta'=1, \bar\zeta=1, \kappa_0=1, \lambda_\rho=2$, and
$q=1$. Here $\Delta\mu>0$ for all
the plots. Nonzero imaginary
part implies propagating modes (see text).}
\label{highfric}
\end{figure}

\begin{figure}[htb]
\includegraphics[height=6cm]{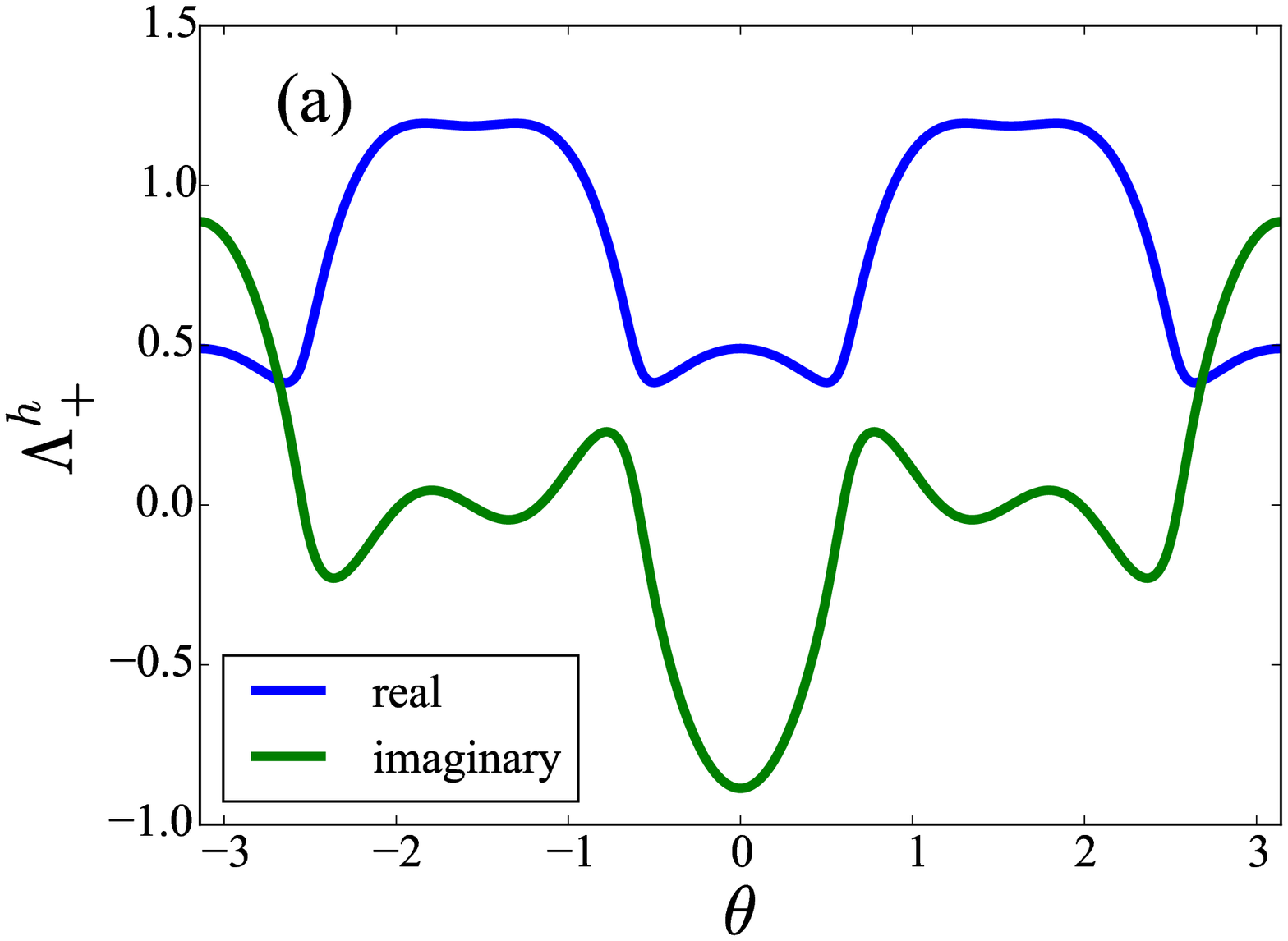},
\includegraphics[height=6cm]{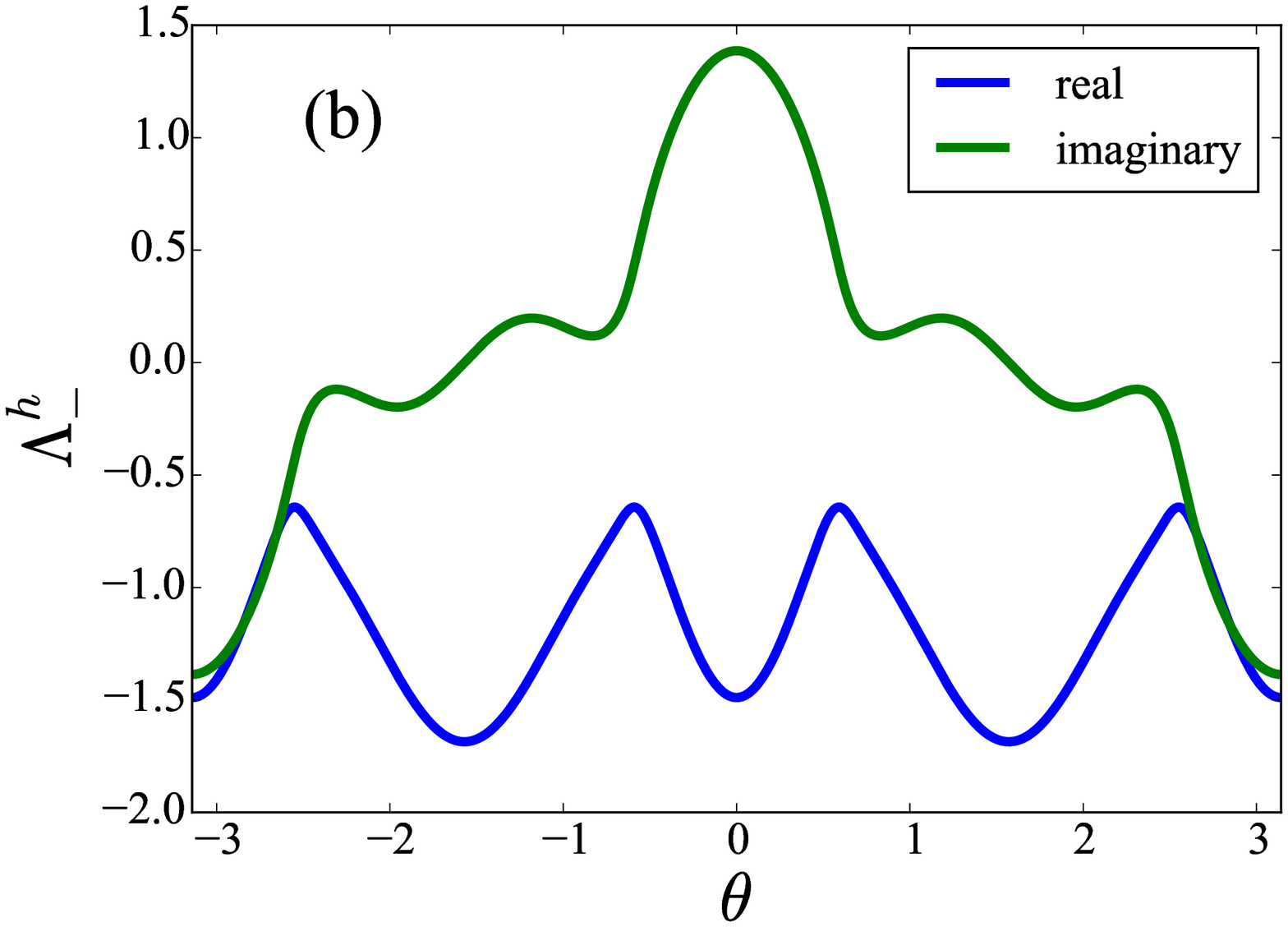}\\
\includegraphics[height=6cm]{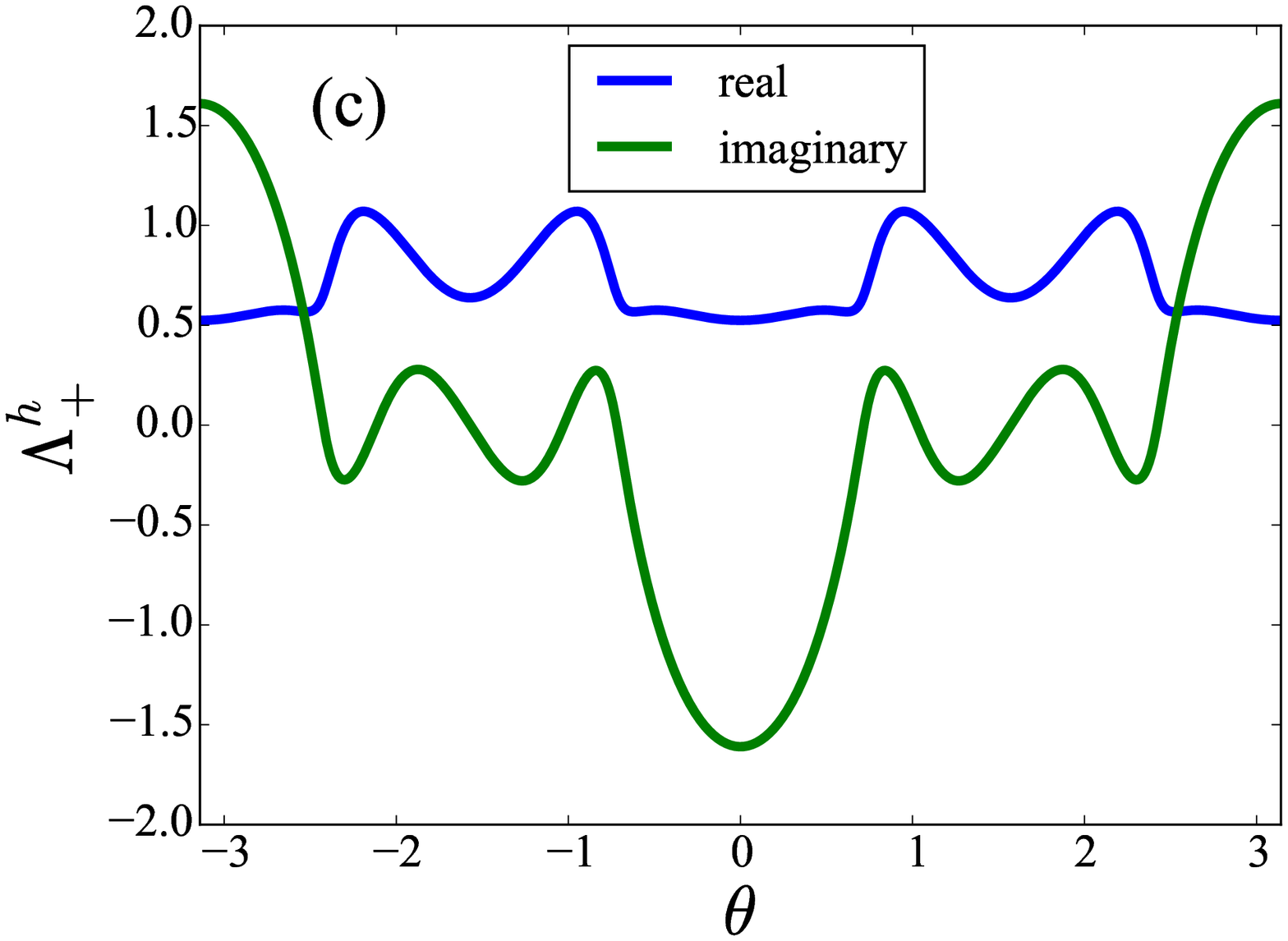},
\includegraphics[height=6cm]{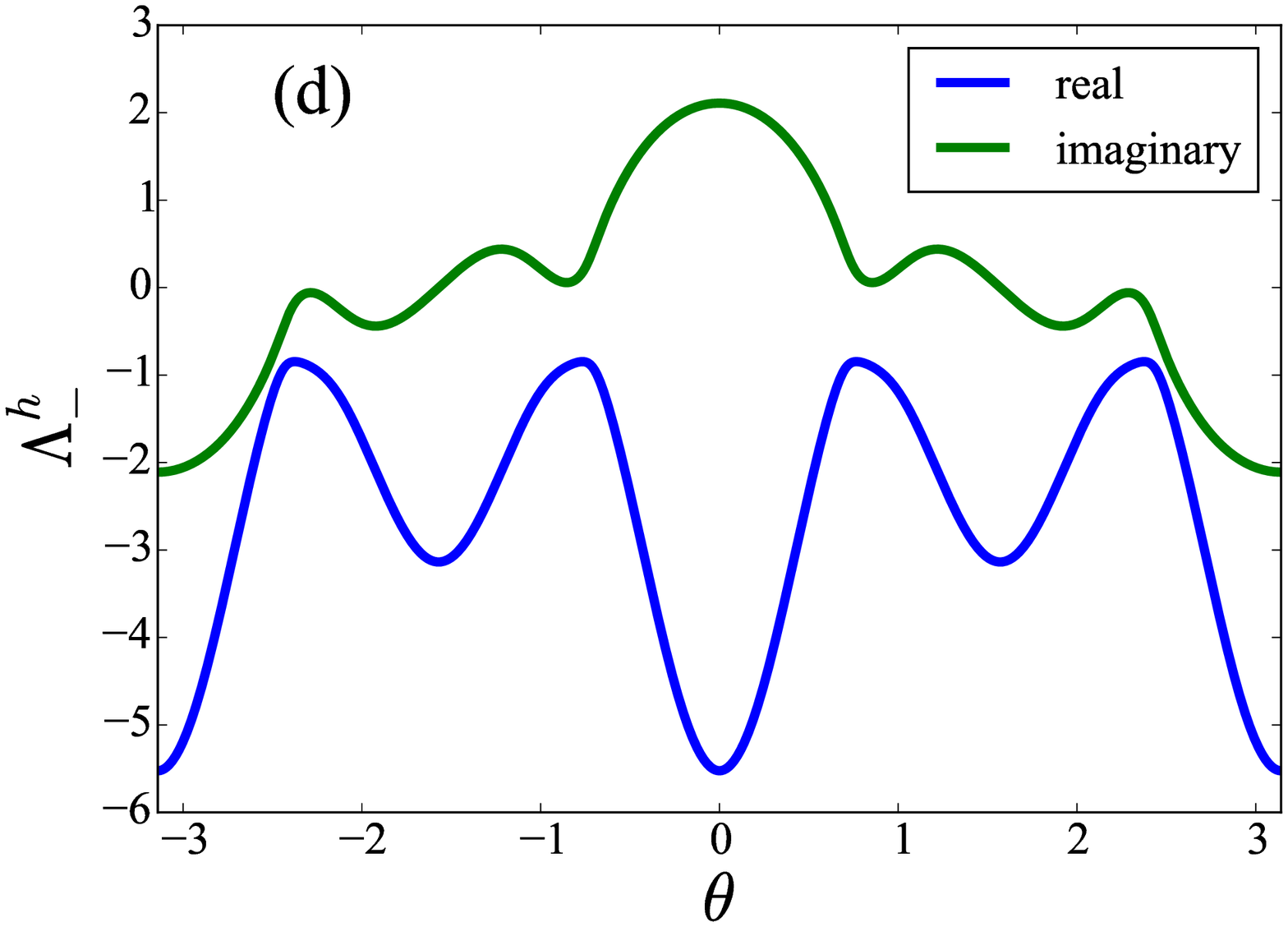}\\
\caption{(Color online) Representative plots of the real (blue line) and imaginary
(green line) parts of (a) eigenmode $\Lambda^h_+$ vs $\theta$ for $\zeta=-1$,
(b) eigenmode $\Lambda^h_-$ vs $\theta$ for $\zeta=-1$,
(c) eigenmode $\Lambda^h_+$ vs $\theta$ for $\zeta=-5$, and
(d) eigenmode $\Lambda^h_-$ vs $\theta$ for $\zeta=-5$
with fixed values of the other parameters; $\lambda_2=1, \kappa_{\rho\rho}=1/2,
\Delta\mu=1, \nu_1=3, \eta'=1, \bar\zeta=1, \kappa_0=1, \lambda_\rho=2$, and
$q=1$. Nonzero imaginary
part implies propagating modes.}
\label{highfriczetaminus}
\end{figure}

\begin{figure}[htb]
\includegraphics[height=6cm]{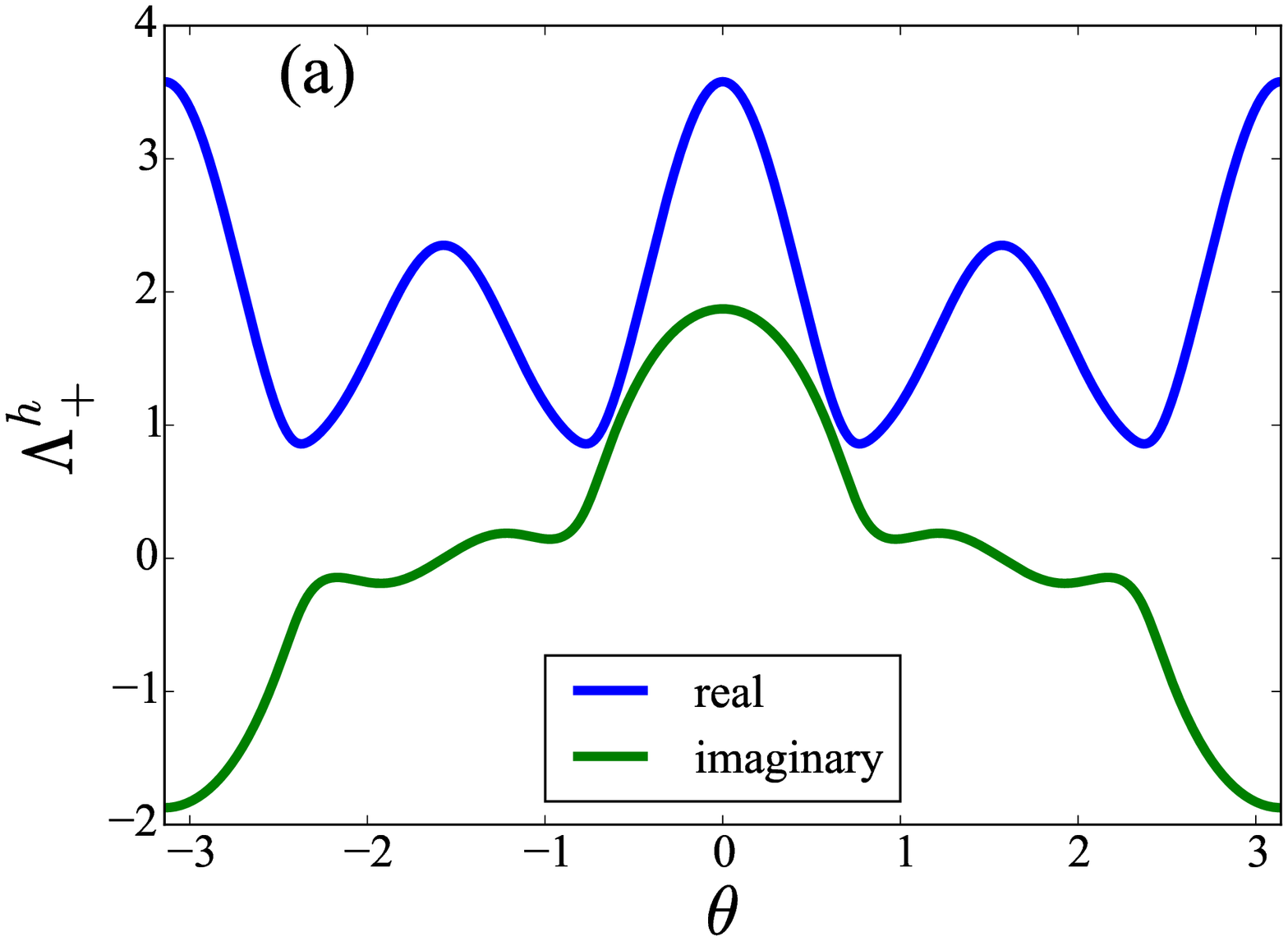},
\includegraphics[height=6cm]{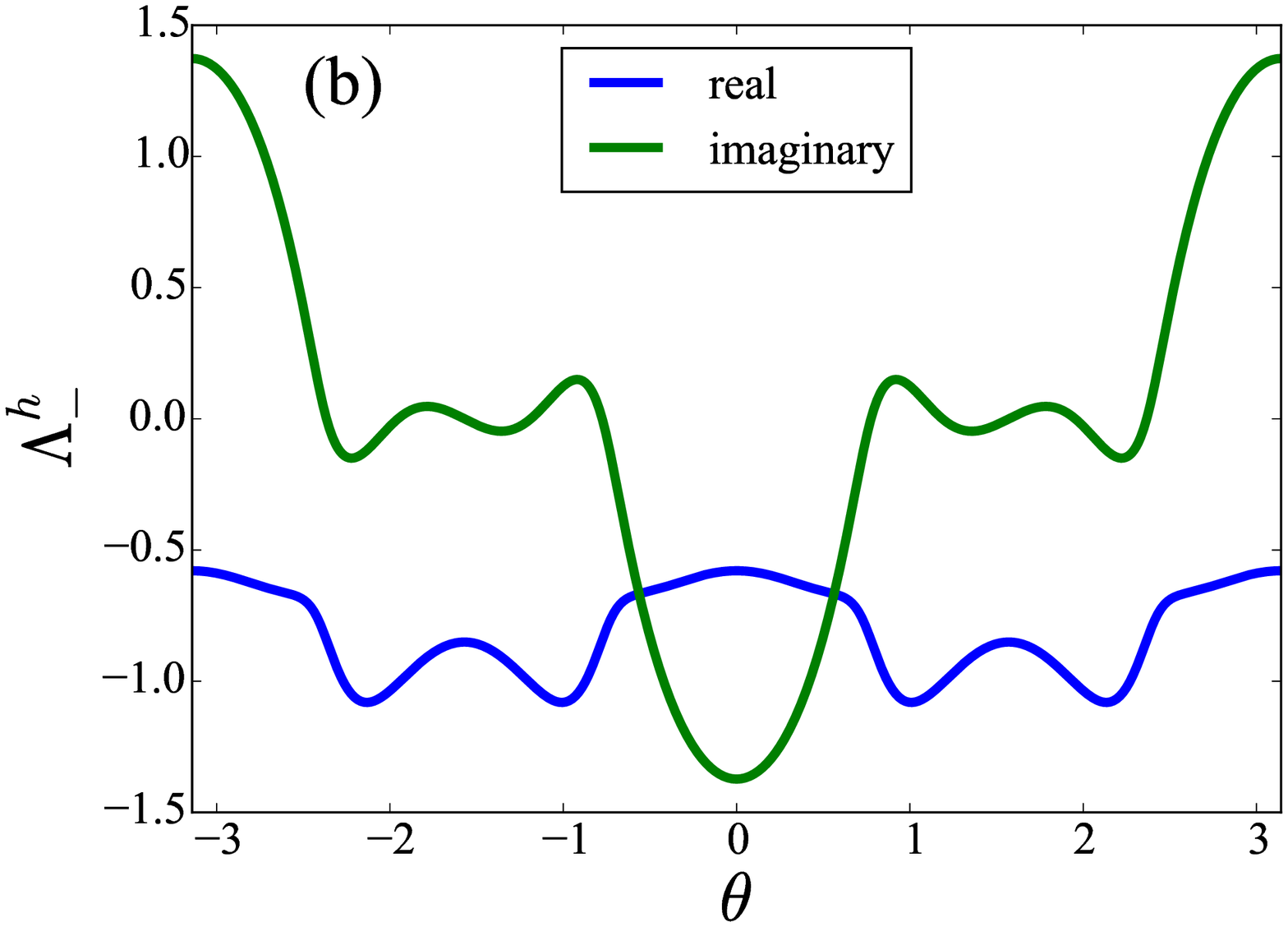}\\
\includegraphics[height=6cm]{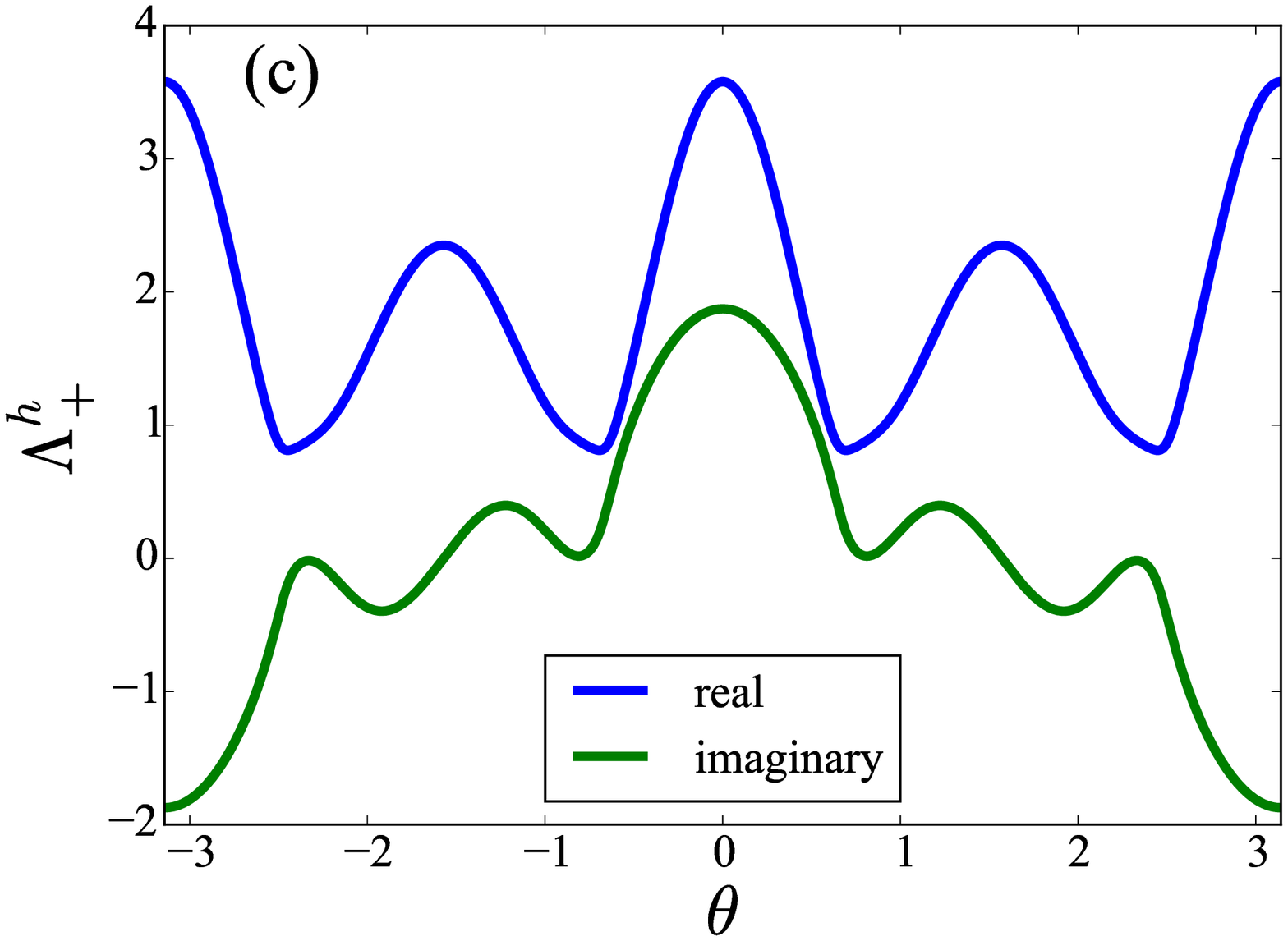},
\includegraphics[height=6cm]{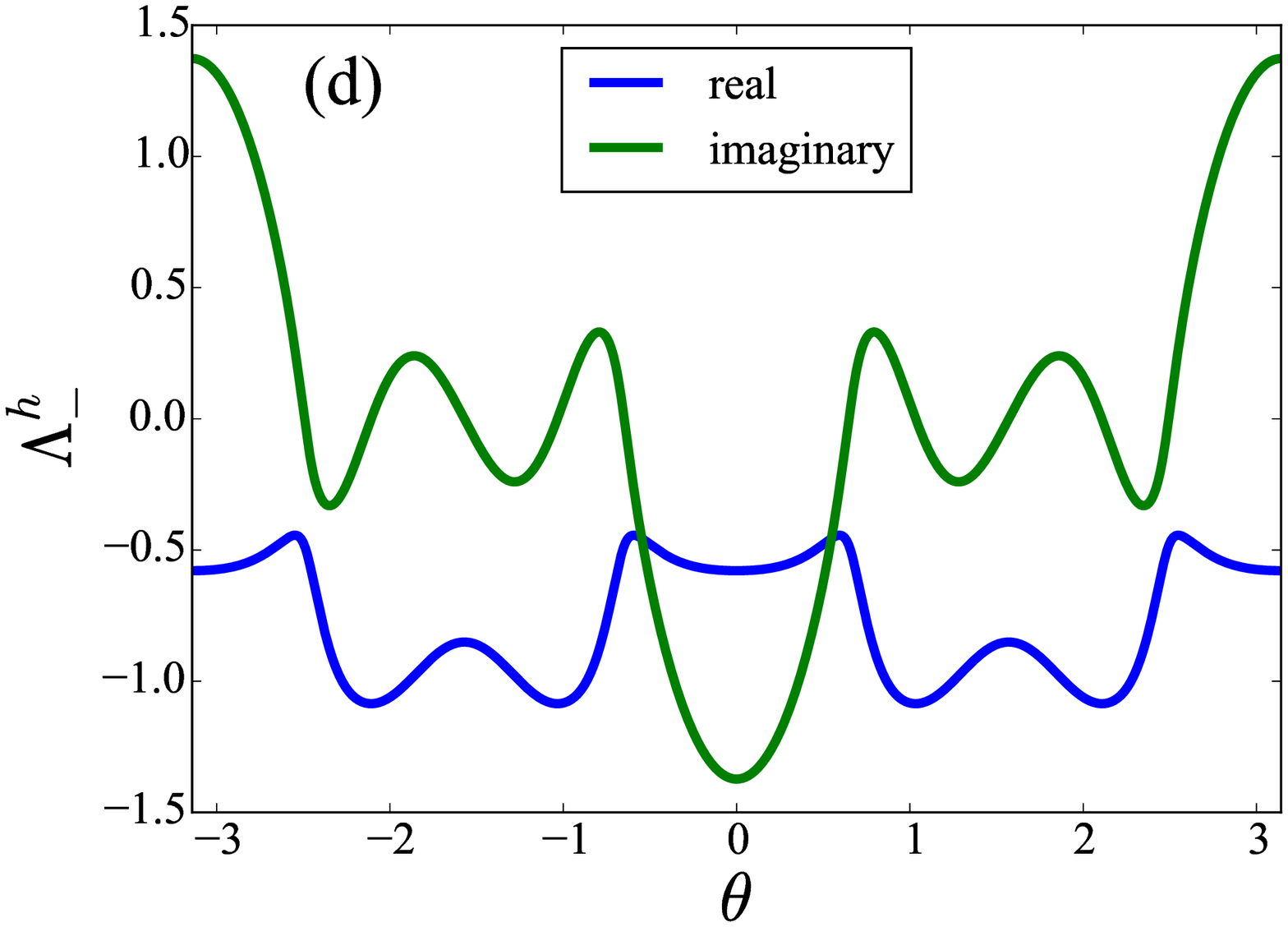}\\
\caption{(Color online) Representative plots of the real (blue line) and imaginary
(green line) parts of (a) eigenmode $\Lambda^h_+$ vs $\theta$ for $\bar\zeta=3$,
(b) eigenmode $\Lambda^h_-$ vs $\theta$ for $\bar\zeta=3$,
(c) eigenmode $\Lambda^h_+$ vs $\theta$ for $\bar\zeta=-3$,
(d) eigenmode $\Lambda^h_-$ vs $\theta$ for $\bar\zeta=-3$,
with fixed values of the other parameters; $\lambda_2=1, \kappa_{\rho\rho}=1/2,
\Delta\mu=1, \nu_1=3, \eta'=1, \zeta=3, \kappa_0=1, \lambda_\rho=2$, and
$q=1$. Nonzero imaginary
part implies propagating modes.}
\label{highfricbarzeta}
\end{figure}

\begin{figure}[htb]
\includegraphics[height=6cm]{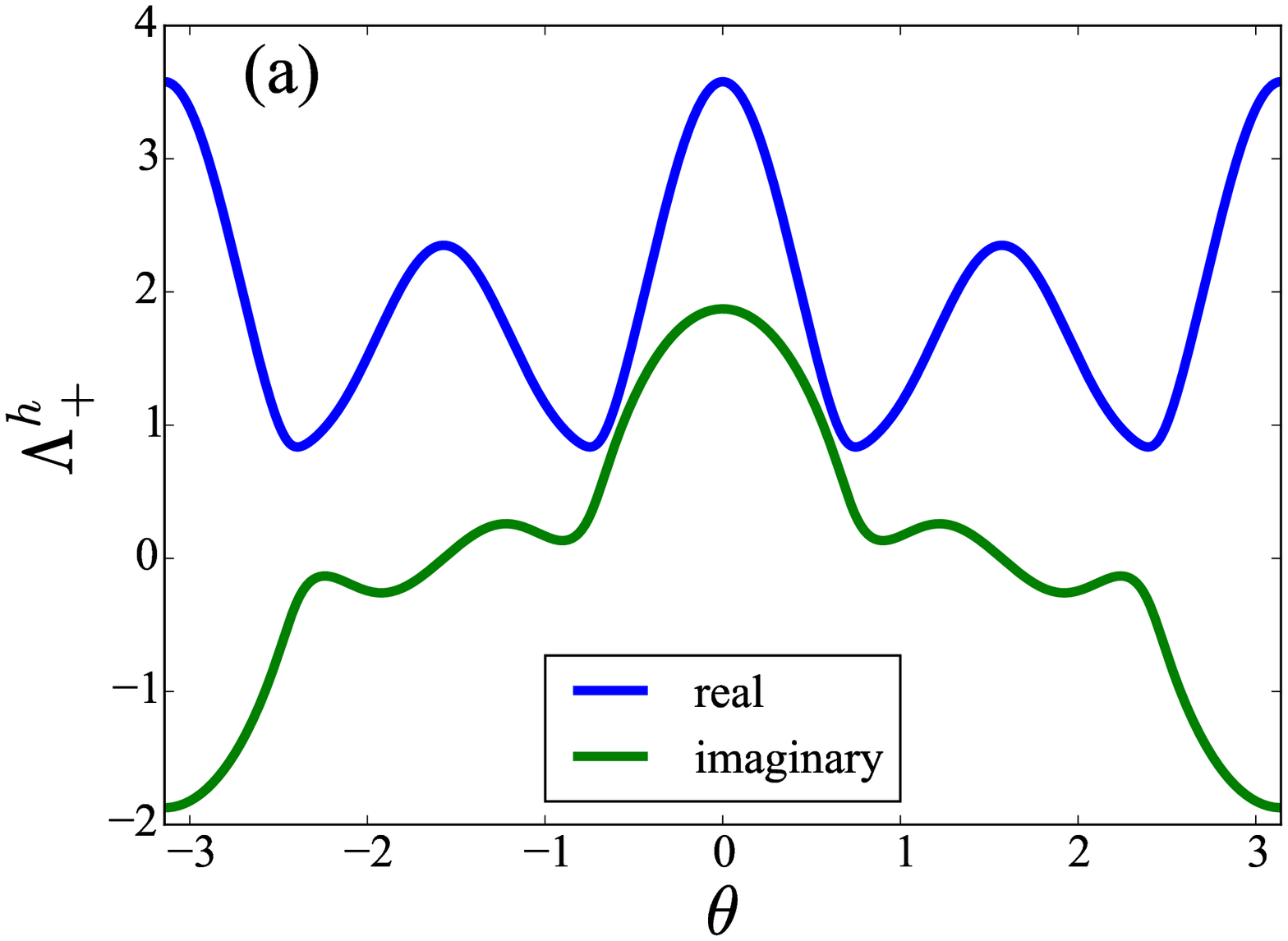},
\includegraphics[height=6cm]{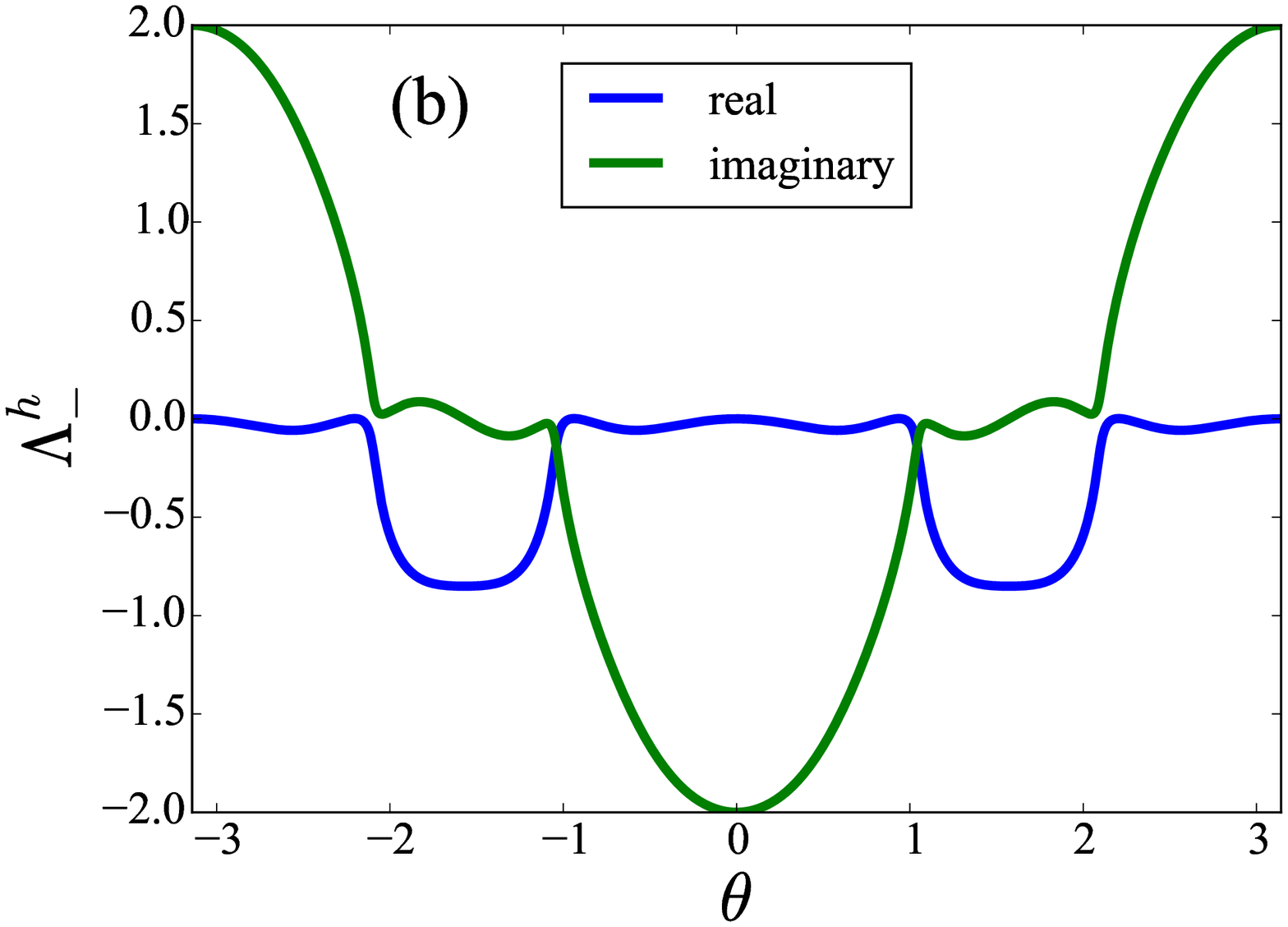}\\
\includegraphics[height=6cm]{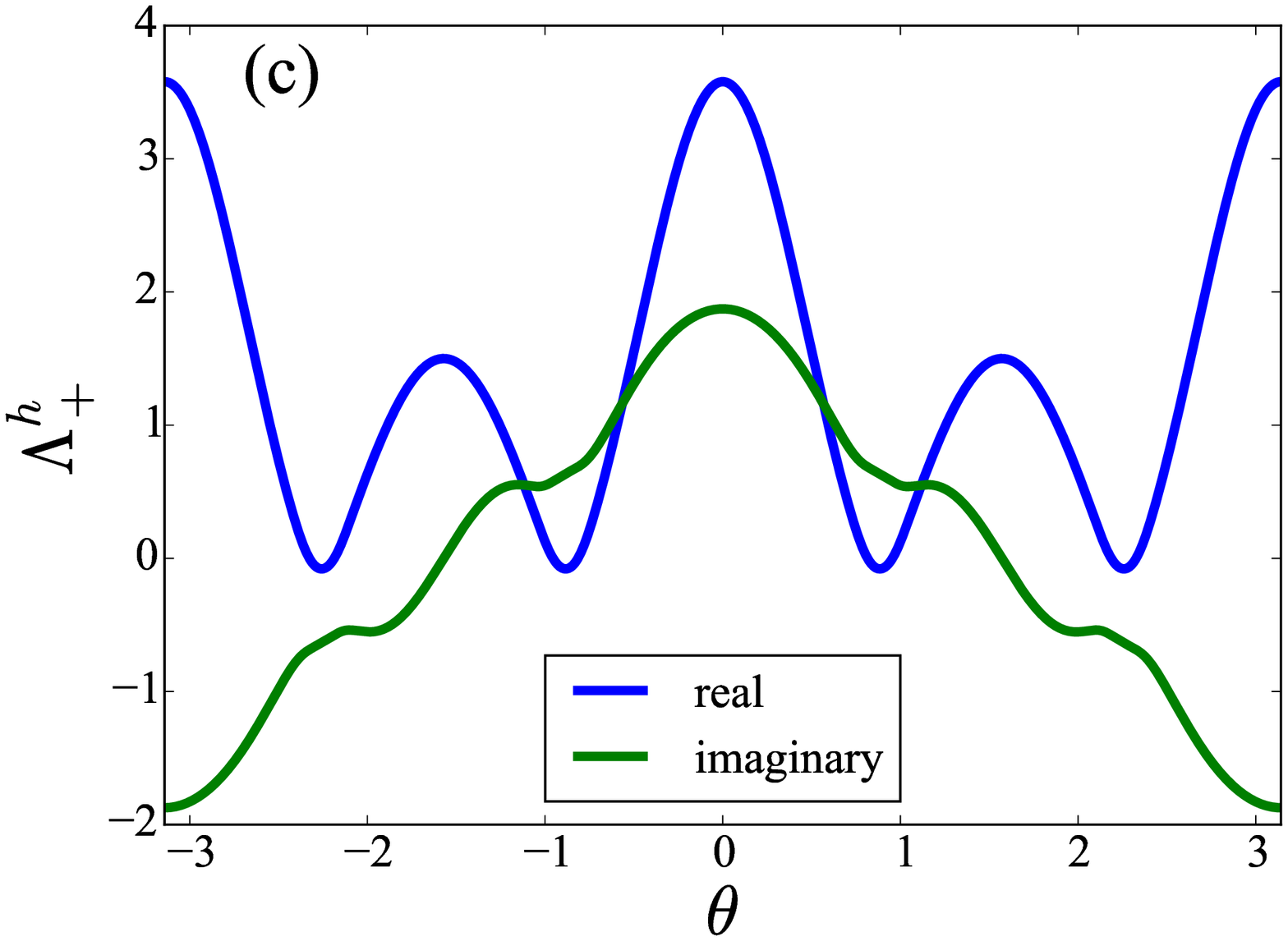},
\includegraphics[height=6cm]{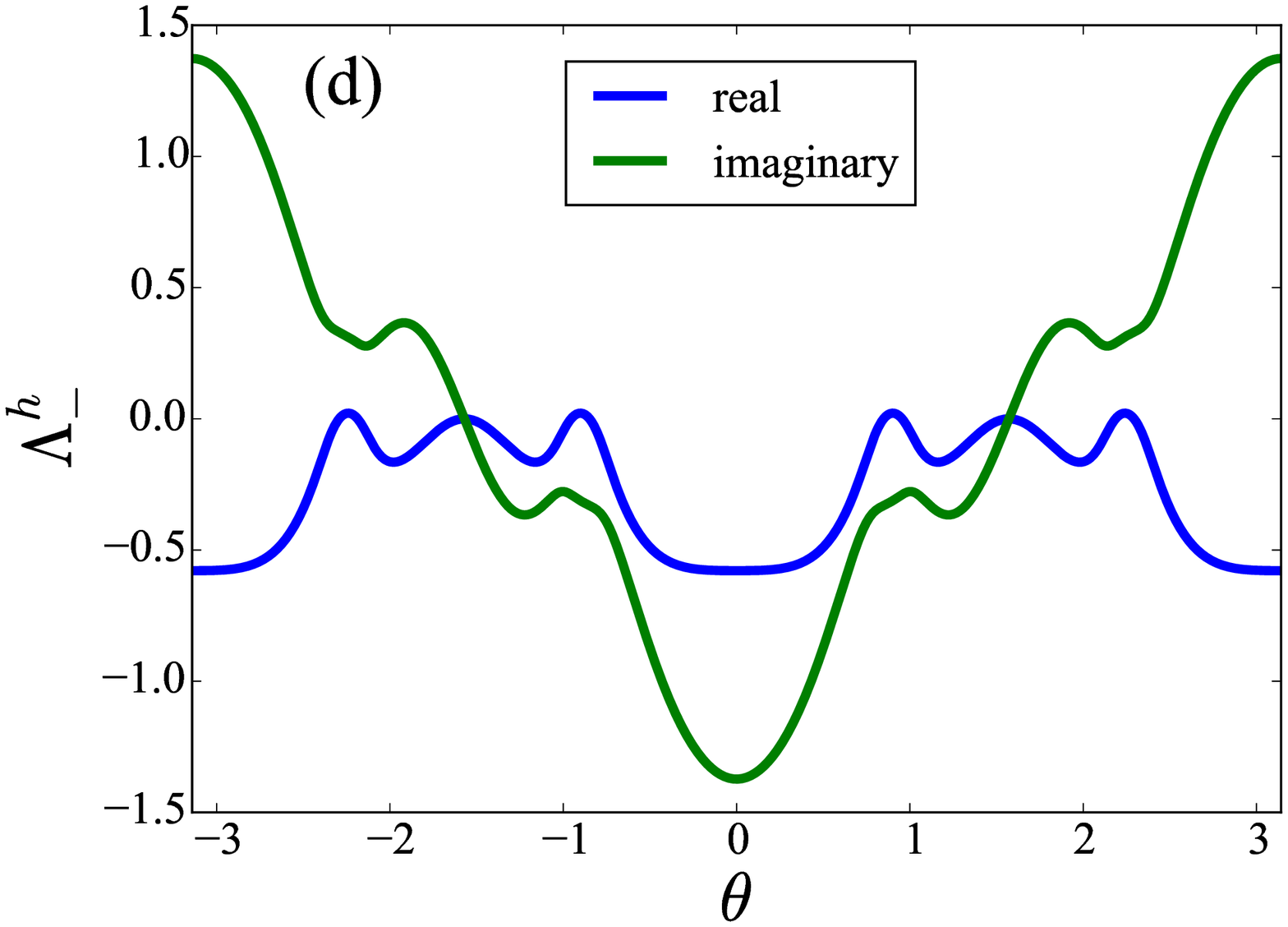}\\
\includegraphics[height=6cm]{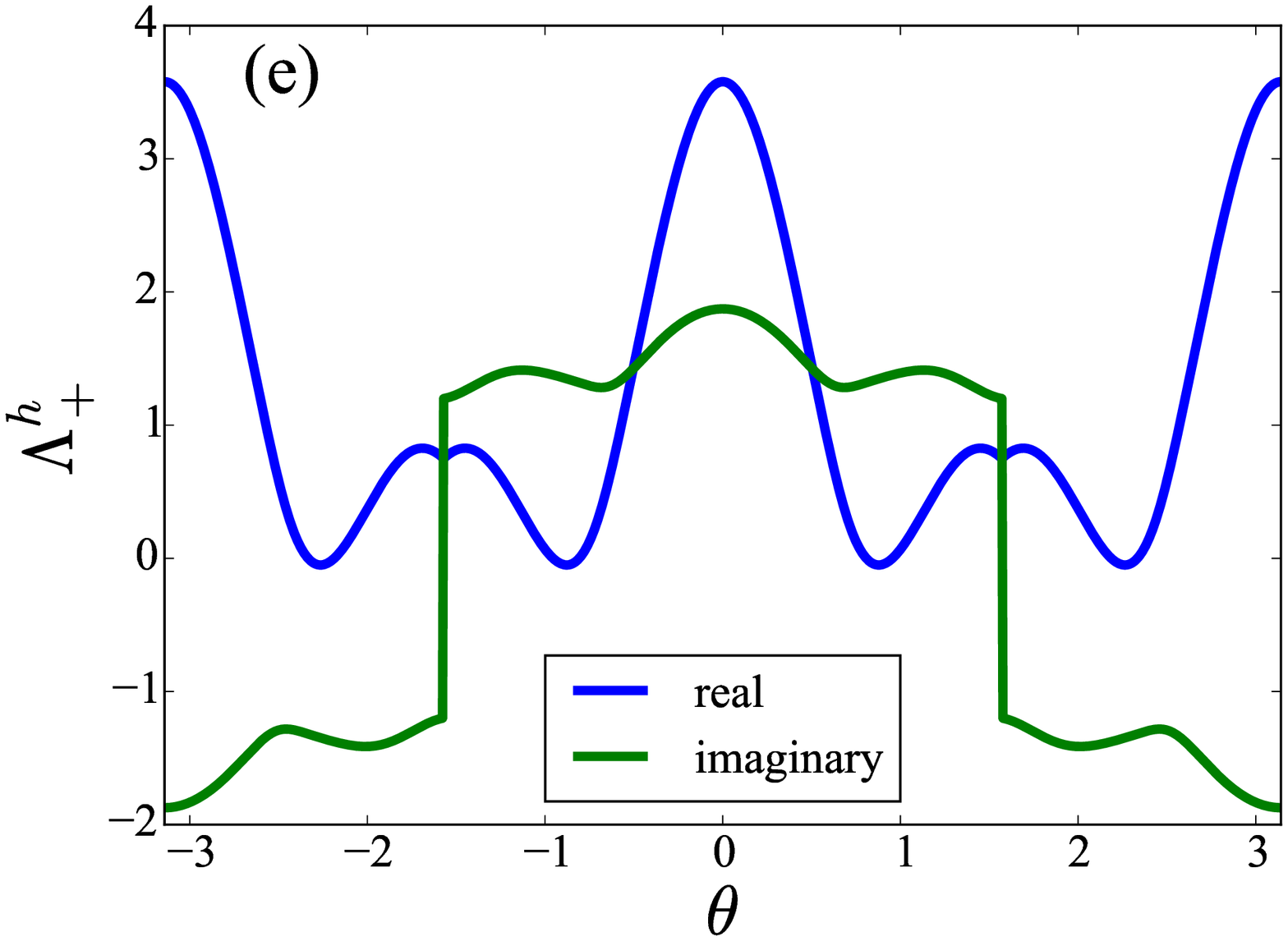},
\includegraphics[height=6cm]{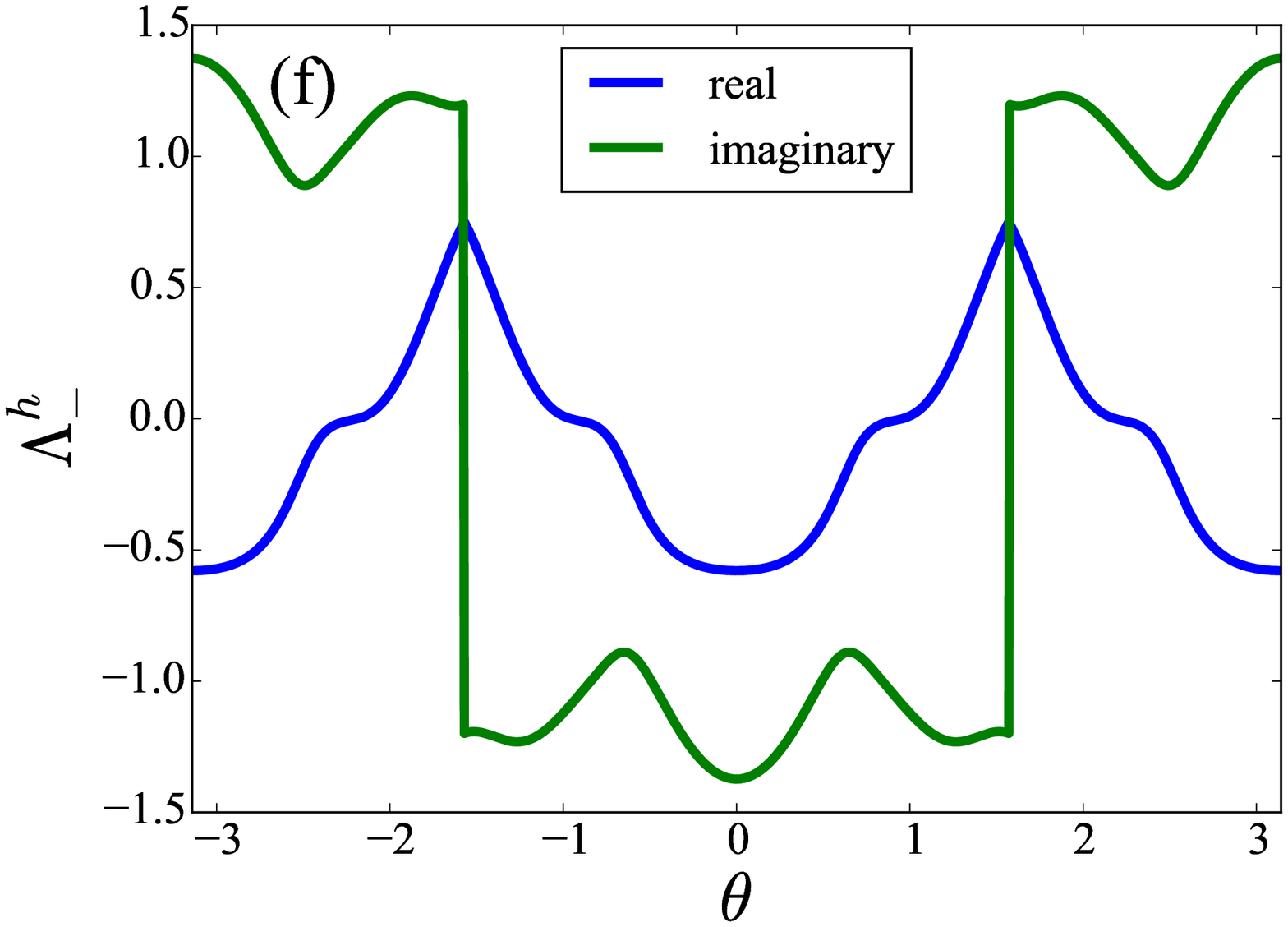}\\
\caption{(Color online) Representative plots of the real (blue line) and imaginary
(green line) parts of (a) eigenmode $\Lambda^h_+$ vs $\theta$ for $\lambda_\rho=2$,
(b) eigenmode $\Lambda^h_-$ vs $\theta$ for $\lambda_\rho=2$,
(c) eigenmode $\Lambda^h_+$ vs $\theta$ for $\lambda_\rho=0$,
(d) eigenmode $\Lambda^h_-$ vs $\theta$ for $\lambda\rho=0$,
(e) eigenmode $\Lambda^h_+$ vs $\theta$ for $\lambda_\rho=-2$,
(f) eigenmode $\Lambda^h_-$ vs $\theta$ for $\lambda\rho=-2$,
with fixed values of the other parameters; $\lambda_2=1, \kappa_{\rho\rho}=1/2,
\Delta\mu=1, \nu_1=3, \eta'=1, \zeta=3, \kappa_0=1, \bar\zeta=1$, and
$q=1$. Nonzero imaginary
part implies propagating modes.}
\label{highfriclambdarho}
\end{figure}

Consider first strong nonequilibrium partial pressure, i.e.,
$\chi/\gamma_0-\bar\lambda\ll \lambda_\rho\Delta\mu$.
The eigenvalues $\Lambda$ of the stability matrix corresponding to
Eqs.~(\ref{pyeq}) and (\ref{rhoeq}) in polar coordinates  ${\bf q}=(q\cos\theta,q\sin\theta)$,
where $\theta$ is the angle between the wavevector ${\bf q}$ and the
ordering direction ($x$-axis), up to the linear order in $q$ are

 \bea \Lambda &=& i{(\lambda_2-\kappa_{\rho\rho})\Delta\mu
\over 2} q\cos\theta + {B\zeta \Delta\mu \over 8\eta'}
q\cos{2\theta} \pm \frac{q\Delta\mu}{2}[
\{i(\lambda_2-\kappa_{\rho\rho}) \cos\theta + {B\zeta
 \over 4\eta'}\cos{2\theta}\}^2 \nonumber \\
&&+ 4\lambda_{\rho}\kappa_0 \sin^2{\theta} + i{B\bar\zeta\kappa_0
\over 2\eta'}\sin\theta\sin{2\theta} + i{B\kappa_{\rho\rho}\zeta \over
\eta'}\cos\theta
\cos{2\theta} \nonumber \\
&&-4\lambda_2\kappa_{\rho\rho}\cos^2\theta]^{1/2}=\Lambda^h_+,\Lambda^h_-. \label{eigen}
\eea
where, $B=(\nu_1+1)\cos^2\theta -(\nu_1-1)\sin^2\theta$.
Clearly, both $\Lambda_+^h,\Lambda_-^h$ scale with $q$ and $\Delta\mu$. Thus,
$\Lambda_+^h,\Lambda_-^h\sim q$, the coefficients of  proportionality are
generally unequal and should in general be complex
functions of $\theta$ (hence anisotropic) and other model parameters. This
linear $q$-dependence is
different from $q$-independent eigenmodes in bulk polar active
fluids (see, e.g., Ref.~\cite{sriram-aditi}) and is a consequence of
the hydrodynamic interactions mediated by the ambient fluid.
In Figs.~\ref{highfric}, representative plots of
$\Lambda_+^h,\Lambda_-^h$ as functions of $\theta$ are shown for two
different values of $\zeta$, namely $\zeta=1$ and $\zeta=5$ for
fixed values of other parameters and $q=1,\Delta\mu>0$. The plots
clearly display a significant change in the amplitude of the real part 
with $\zeta$ in one of the eigenvalues and in the amplitude of the 
imaginary part with $\zeta$ in the other one. It may thus be concluded that 
the amplitude of the unstable mode and as well as the propagating mode changes with 
change in the magnitude of $\zeta$ or the active stress coefficient. 
Figures~\ref{highfriczetaminus} 
show plots of the eigenvalues as function of $\theta$ for negative values of $\zeta$,
keeping all other parameters fixed, which again shows the change in
amplitude of the real and the imaginary parts in the two different 
eigenvalues, with change in value of $|\zeta|$.
Figures~\ref{highfricbarzeta} compare the eigenvalues for different
signatures of $\bar\zeta$ (the coefficient of the small fluctuations
of the active stress), keeping all other parameters fixed. From the
plots it is quite clear that the dependence of $\Lambda_+^h$ and
$\Lambda_-^h$ on $\bar\zeta$ is very weak. In
Figs.~\ref{highfriclambdarho} the plots of $\Lambda_+^h$ and
$\Lambda_-^h$ are shown for different signs and values of
$\lambda_\rho$, other parameters kept constant. The plots bring out
the changes in the imaginary parts. Furthermore, although the real
part of $\Lambda_+^h$ is always found to be positive (for these
choices of the parameters) and signifies instability in the system,
the real part of $\Lambda_-^h$ shows a transition from typically
negative values for $\lambda_\rho>0$ to positive values for
$\lambda_\rho<0$. This suggests a very strong dependence of the
eigenmodes on the value and signature of $\lambda_\rho$ or the active osmotic 
pressure coefficient. 

Consider in detail the limit $B\rightarrow 0$, or,
\beq
\tan^2\theta= {\nu_1+1 \over \nu_1-1}\equiv\tan^2\theta_0,
 \eeq yielding $\theta=\pm\theta_0,\pm (\theta_0+\pi)$, such that $B(\theta_0)=0$. The
eigenvalues in this case are \bea \Lambda(\theta_0)=
i{(\lambda_2-\kappa_{\rho\rho})\Delta\mu \over 2} q\cos\theta_0 \pm
{\Delta\mu q \over 2}[-(\lambda_2+\kappa_{\rho\rho})^2\cos^2\theta_0
+4\lambda_\rho \kappa_0\sin^2\theta_0]^{1/2}. \label{eigenthetanot}
\eea It is evident from Eq.~(\ref{eigenthetanot}) that for
$\lambda_\rho \kappa_0<0$, $\Lambda(\theta_0)$ is fully imaginary
i.e., two propagating modes, which are oppositely moving, are
present in the system with an anisotropic $q$-independent wave
speed. Thus, $\theta_0$ gives the direction in the plane along
which small fluctuations propagate without growth or decay at
$O(q)$. On the other hand, for $\lambda_\rho \kappa_0>0$ and
$|4\lambda_\rho\kappa_0\sin^2\theta_0|>|(\lambda_2
+\kappa_{\rho\rho})^2\cos^2\theta_0|$, $\Lambda(\theta_0)$ has a
real part in addition to propagating modes. The real part comes from
\beq \pm [-(\lambda_2+\kappa_{\rho\rho})^2\cos^2\theta_0
+4\lambda_\rho \kappa_0\sin^2\theta_0]^{1/2}. \label{combi}\eeq
Evidently, the real part displays instability for both signs of
$\Delta\mu$  in this case, with anisotropic decay/growth rates which
scale with $q$. Thus, unlike the case with
$\lambda_\rho\kappa_0<0$, there is no particular significance of the
angle $\theta_0$ here.

\begin{figure}[htb]
\includegraphics[height=6cm]{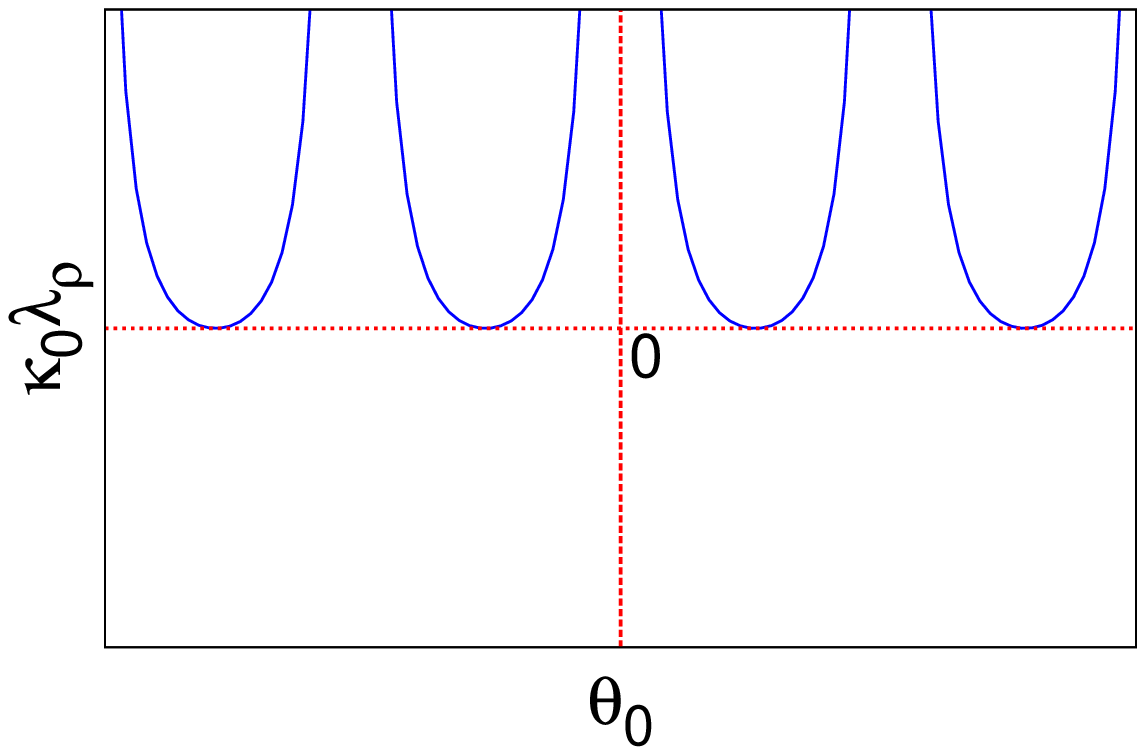}
\caption{(color online) A plot of $\kappa_0\lambda_\rho$ vs $\theta_0$ for some chosen
values of $\kappa_{\rho\rho}$ and $\lambda_2$ is shown. The regions inside the upward
parabolas indicate the presence of moving instabilities and all
other regions outside have propagating modes without damping or growth at $O(q)$ (see text).}
\label{gnuinterfric}
\end{figure}

A schematic plot of $\kappa_0\lambda_\rho$ vs $\theta$ for chosen
values of $\kappa_{\rho\rho}$ and $\lambda_2$ is shown in
Fig.~(\ref{gnuinterfric}), clearly indicating the unstable regions
and propagating modes. That the expression (\ref{combi})
determines the stability at $\theta=\theta_0$ be understood
heuristically as follows. If all other parameters are set to to
zero, the combination $\lambda_2 +\kappa_{\rho\rho}$ gives the
relative speed of propagations of the fluctuations of $p_y$ and
$\rho$ in the linearised theory, with $p_y$ and $\rho$ being
decoupled from each other. On the other hand the product
$\kappa_0\lambda_\rho$ controls the wavespeed or the growth rate of
the fluctuations in the linearised coupled system of $p_y$ and
$\rho$, depending on its sign, with all other parameters set to
zero. Thus, in a situation where all the above four parameters are
nonzero, it is generally expected that $\lambda_2
+\kappa_{\rho\rho}$ has the effect of stabilising the instabilities
due to $\kappa_0\lambda_\rho$ (assuming it has the sign that
corresponds to instability), by allowing reduction of local
inhomogeneities to disperse
 by means of wave
propagation.

\begin{figure}[htb]
\includegraphics[height=6cm]{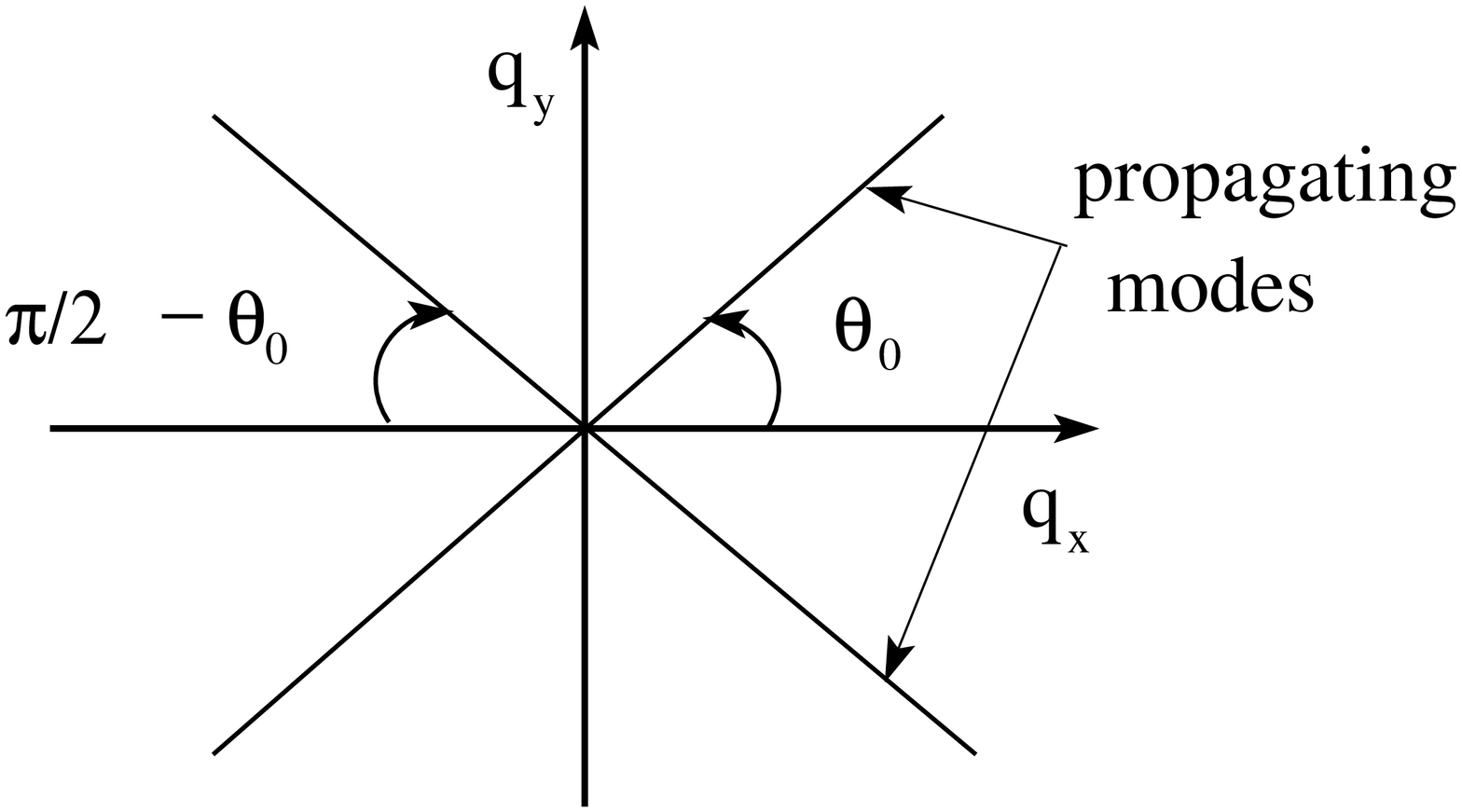}
\caption{Schematic diagram  ($\kappa_0\lambda_\rho <0$ or $\Delta<0$) displaying
the special angular directions
given by $\theta=\theta_0$ in the plane along which there are only
propagating modes up to $O(q)$ (see text).}
\label{theta0hf}
\end{figure}

Consider the case when there are only propagating modes at $O(q)$ at, say,
$\theta=\theta_0$. Now assume $\theta$ very close to $\theta_0$; we
write $\theta=\theta_0 +\delta\theta$, where $\delta\theta$ is
very small. In that case $B\approx -2\nu_1\sin{2\theta_0}
\delta\theta$, up to order $O(\delta\theta)$. The eigenvalues
corresponding to $\theta=\theta_0+\delta\theta$ are given by \bea
\Lambda(\theta_0+\delta\theta)=\Lambda(\theta_0)-{\zeta\Delta\mu\nu_1
\over 4\eta'} q\sin{2\theta_0}\cos{2\theta_0}\delta\theta +
iO(\delta\theta).\label{thetaimag11}
 \eea
  Noting that $\Lambda(\theta_0)$ is fully imaginary, (\ref{thetaimag11})
  shows that $\Lambda (\theta_0+\delta\theta)$ has real parts, whose signs depend on
$\Delta\mu$ for a given $\delta\theta$.  Thus, we conclude that the
system shows instability for either sign of $\Delta\mu$ along with
generic propagating modes with an anisotropic wave speed
proportional to $\Delta\mu$. 
 Considering $\Lambda$ in the $(q,\theta)$ plane, we thus notice that
 there are special directions given by $\theta=\pm\theta_0,\pm (\theta_0+\pi)$
 along which (small) perturbations move as waves without any growth or damping
 (to the linear order in $q$), provided
 $\Delta=4\lambda_\rho\kappa_0\sin^2\theta_0- (\lambda_2+\kappa_{\rho\rho})^2
 \cos^2\theta_0 <0$ is satisfied; see Fig.~\ref{theta0hf}. Additional values of $\theta$ for which for
 which the real parts of $\Lambda_+^h$ or $\Lambda_-^h$ vanish may be
 found from (\ref{eigen}). However, both the real parts will not vanish
 simultaneously at these angles; see Fig.~\ref{highfric}. Along all other
 directions, at least one of $\Lambda_{\pm}^h$ should have a real part,
 and hence perturbations will grow/decay and move.
 If $\Delta >0$, there are no special directions with only propagating modes.
 Since, $p_y({\bf x},t)$ and $\rho({\bf
x},t)$ depend on $p_y({\bf q},t)$ and $\rho({\bf q},t)$ for all $\bf
q$, hence,  the two eigenmodes for all $\bf q$, $p_y({\bf x},t)$ and
$\rho({\bf x},t)$ show generic moving instabilities at $O(q)$ for
both signs of $\Delta\mu$ for arbitrary choice for the active
coefficients. It is also clear that at $O(q)$, the system can be
stable only if $\zeta=0=\overline\zeta$ and $\lambda_\rho\kappa_0 <0$. Thus,
the active stresses clearly distablise the system. Of course, at
higher order in $q$, the system will be stabilised by large enough $D$ or $\gamma_{\rho\rho}$.


It is useful to analyse the stability of
the system for some particular values of $\theta$.
First we start with $\theta=0$. In this limit the eigenvalues are
given by \bea \Lambda(\theta=0) = i\lambda_2\Delta\mu q +
{(\nu_1+1)\zeta \Delta\mu \over 4\eta'}q, -i\Delta\mu
\kappa_{\rho\rho}q. \label{eigen0theta} \eea Thus there are two
modes; one is purely imaginary and hence just a propagating mode,
the other has both real and imaginary parts. The sign of the real
part is determined by $\Delta\mu$. Thus this eigenvalue is moving
and either growing (unstable) or decaying (stable) in time, respectively,
when $(\nu_1+1)\zeta\Delta\mu >0$, or, $<0$.

For $\theta={\pi \over 2}$, the stability eigenvalues are given by
\bea
\Lambda(\theta={\pi \over 2}) = {(\nu_1 -1)\zeta\Delta\mu \over 8\eta'}q
\pm {\Delta\mu q \over 2}\left[\left({(\nu_1 -1)\zeta  \over 4\eta'}
\right)^2 + 4\lambda_{\rho}\kappa_0 \right]^{1/2}.
\label{eigen90theta}
\eea
From Eq.~(\ref{eigen90theta}), we note that for $\lambda_\rho\kappa_0>0$,
the system is unstable for both $\Delta\mu>0$
and $\Delta\mu<0$. Next,
for $\lambda_\rho
\kappa_0<0$ and $(\nu_1-1)\zeta >0$,
\begin{itemize}
 \item If $|4\lambda_\rho\kappa_0|> {(\nu_1-1)^2\zeta^2 \over 4\eta'^2}$ and
 $\Delta\mu>0$, the modes are unstable and oppositely moving.
\item However, when $|4\lambda_\rho\kappa_0 |< {(\nu_1-1)^2\zeta^2 \over 4\eta'^2}$
with $\Delta\mu>0$, both the modes are unstable.
There are no propagating waves.
\end{itemize}

In the special case with $\bar\zeta=0=\lambda_{\rho}$ in Eq.~(\ref{eigen}), i.e., if we
ignore the density  dependences of the active coefficients, the
eigenvalues of the stability matrix take a simpler form
 \bea
\Lambda(\bar\zeta=0=\lambda_\rho)=i\lambda_2\Delta\mu q\cos\theta +
{B\zeta\Delta\mu \over 4\eta'}q\cos{2\theta},
-i\kappa_{\rho\rho}\Delta\mu q\cos\theta, \eea which indicates the
presence of propagating modes and instability for both signs of
$\Delta\mu$ above or below $\theta={\pi \over 4}$.

Now briefly consider the instabilities with $|\chi/\gamma_0 -\bar\lambda|\gg
|\lambda_\rho\Delta\mu|$ (weak nonequilibrium partial pressure):
Neglecting $\lambda_\rho\Delta\mu$ in comparison with
$\chi/\gamma_0-\bar\lambda$, the eigenvalues $\Lambda$ are given by
\begin{eqnarray}
\Lambda&=&\frac{\Delta\mu q}{2}[-i\kappa_{\rho\rho}\cos\theta
+\frac{\zeta B}{4\eta'}\cos{2\theta}+i\lambda_2q\cos\theta]\pm\frac{q}{2}\{[-\frac{\zeta
B\cos{2\theta}}{4\eta' }-i\lambda_2
\cos\theta+i\kappa_{\rho\rho}\cos\theta]^2\Delta\mu^2\nonumber \\ &&
+4[\frac{i\Delta\mu}{4\eta'}\overline\zeta
B\sin^2\theta\cos\theta+(\frac{\chi}{\gamma_0}-\bar\lambda)\sin^2\theta]\kappa_0\Delta\mu
+4i\kappa_{\rho\rho}\Delta\mu^2\cos\theta[i\lambda_2 \cos\theta
+\frac{\zeta B}{4\eta'}\cos{2\theta}]\}^{1/2}.
\end{eqnarray}
Thus, $\Lambda$ are no longer homogeneous
functions of $\Delta\mu$. In order to progress further, assume a
"small" $\Delta\mu$. Then, in an expansion in powers
of $\Delta\mu$, we obtain to the lowest order in $q$ and $\Delta\mu$
\bea
\Lambda &=& \pm \sqrt{(\frac{\chi}{\gamma_0}-\bar\lambda)\kappa_0\Delta\mu}
q\sin\theta,\;\;{(\frac{\chi}{\gamma_0}-\bar\lambda)\kappa_0\Delta\mu} >0,\\
\Lambda&=& \pm iq\sin\theta
\sqrt{|(\frac{\chi}{\gamma_0}-\bar\lambda)\kappa_0\Delta\mu|},\;\;
{(\frac{\chi}{\gamma_0}-\bar\lambda)\kappa_0\Delta\mu}
<0. \eea Thus, in the former case, we find instabilities for either
sign of $\Delta\mu$, where as in the second case, we find oppositely
moving propagating waves.

\subsection{Intermediate friction}\label{interlin}

%

\begin{figure}[htb]
\includegraphics[height=6cm]{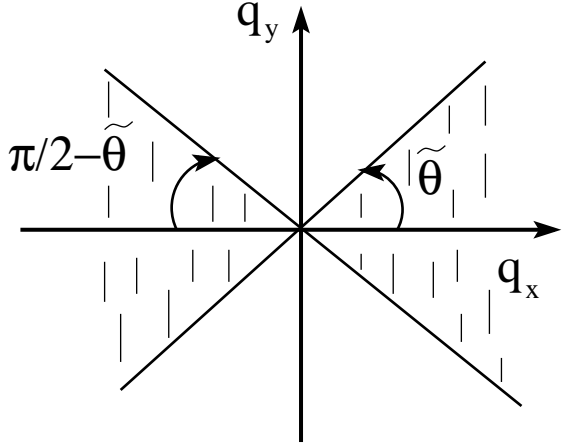}\\
\caption{Schematic diagram ($\kappa_0\lambda_\rho>0$) depicting the regions
of directions in the plane (shaded region) in which only propagating modes exist at $O(q)$.
Outside of the region, the system is unstable at $O(q)$. In contrast, pure
propagating modes without any damping or gworth are found for $l_s\rightarrow 0$
 ($\Gamma\rightarrow\infty$) only along four special
lines for $\kappa_0\lambda_\rho <0$ (see text; see also Fig.~\ref{theta0hf}).}
 \label{eigenint}
 \end{figure}

We again consider $|\chi/\gamma_0-\bar\lambda| \ll |\lambda_\rho\Delta\mu|$ first.
To the lowest order (linear
order) in ${\bf q}$, the eigenvalues of the linear stability matrix are
\bea \Lambda &=& -{i\Delta\mu
q \over 2}(\kappa_{\rho\rho}-\lambda_2)\cos\theta \pm {iq\Delta\mu
\over 2}\left[\cos^2\theta (\kappa_{\rho\rho}+\lambda_2)^2 -
4\kappa_0\lambda_\rho\sin^2 \theta\right]^{1/2} \nonumber \\
&=& \Lambda_+,\Lambda_-,    \label{eigenweak}
\eea
which are {\em independent} of $\Gamma$.
We can make the following general conclusions about the mode structures
from (\ref{eigenweak}). First of all, none of the active stress
coefficients $\zeta$ and $\overline\zeta$  appear in (\ref{eigenweak}).
Thus the active stress is {\em irrelevant} in
the dynamics to the linear order in $\bf q$. The dynamics at this
order in $q$ is controlled by the remaining active coefficients,
{\em viz},  $\kappa_0,\kappa_{\rho\rho},\lambda_2$ and
$\lambda_\rho$. This is clearly in contrast to the situation with large
$\Gamma$. Secondly, if $\kappa_0 \lambda_\rho<0$, then the
discriminant is positive for all values of $\theta$ to the linear order in $q$. Then only
propagating modes will be present for all values
 of $\theta$. As before, there should be two oppositely moving propagating modes
with the speed of wave being anisotropic and proportional to $\Delta\mu$.
If on the other hand $\kappa_0 \lambda_\rho
>0$, the discriminant in (\ref{eigenweak}) is negative for all magnitudes of $\kappa_0
\lambda_\rho\neq 0$ at least at $\theta=\pi/2$, giving rise to instability
in the system for both signs of $\Delta\mu$. These instabilities are
moving in the opposite directions. In general, for any value of
$\theta$ satisfying $\cos^2\theta (\kappa_{\rho\rho} +\lambda_2)^2-
4\kappa_0 \lambda_\rho\sin^2 \theta >0$ and for both signs of
$\Delta\mu$, both the modes are propagating without damping (or growth).
Thus, any perturbation in a region of the polar plane satisfying the
above condition moves without any growth or decay in the amplitude
(up to $O(q)$). Else, in the remaining region of the polar plane, one
of the modes is unstable and the other stable. The
speed of the moving stabilities are unsurprisingly anisotropic and
proportional to $\Delta\mu$. The above consideration for $\kappa_0\lambda_\rho>0$
allows us to define an angle $\tilde\theta$ such that
\beq
 \cos^2\tilde\theta (\kappa_{\rho\rho} +\lambda_2)^2-
4\kappa_0 \lambda_\rho\sin^2 \tilde\theta =0.
 \eeq
 Then, for $\kappa_0\lambda_\rho>0$ in the shaded region in Fig.~\ref{eigenint}
 characterised by $\tilde\theta$ there are only propagating waves at
 $O(q)$, outside of this region, the system is linearly unstable at $O(q)$.

\begin{figure}[htb]
\includegraphics[height=5cm]{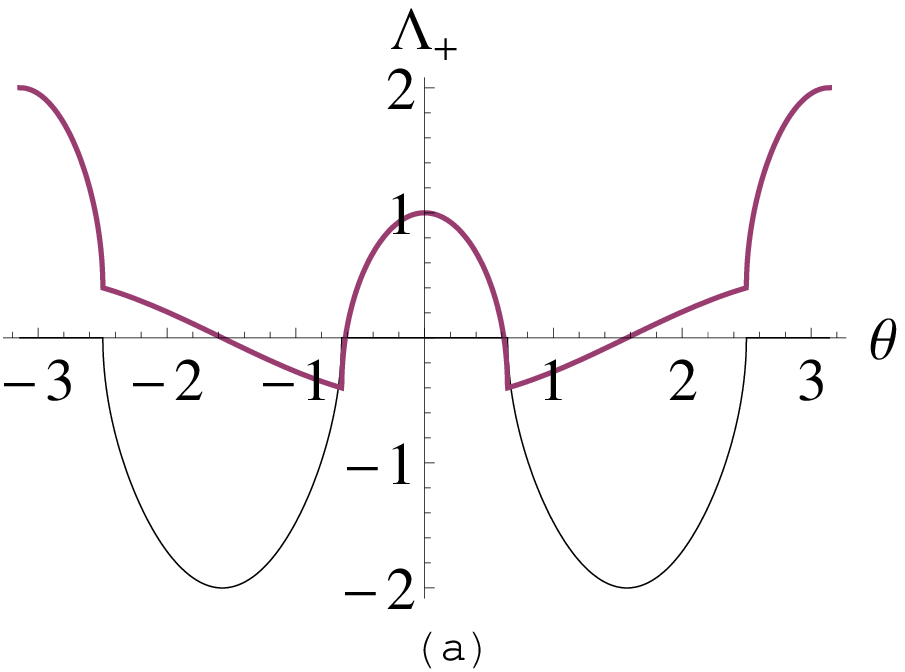}, \includegraphics[height=5cm]{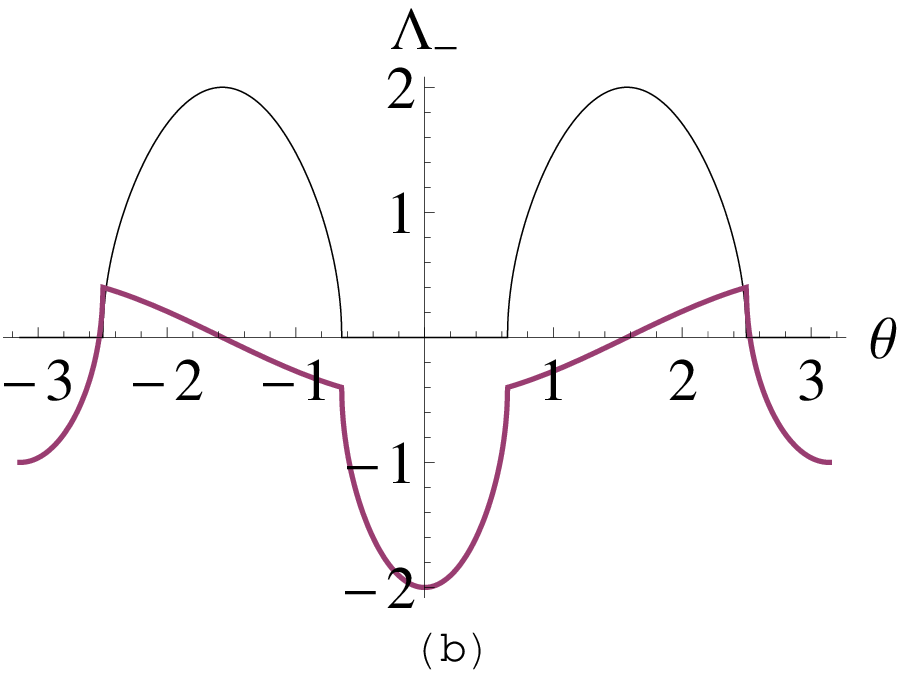} \\
\includegraphics[height=5cm]{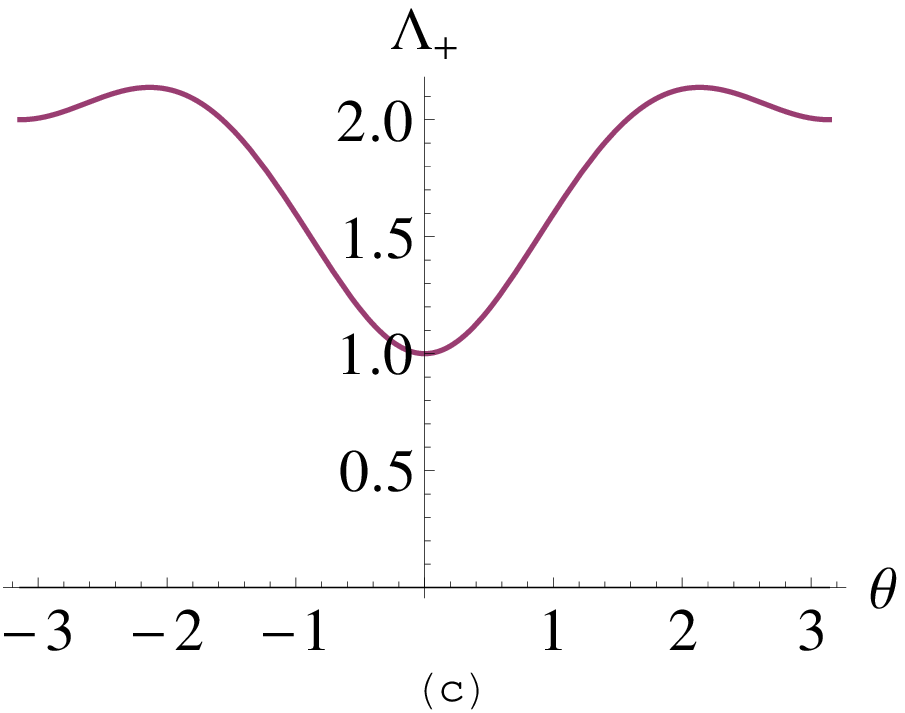}, \includegraphics[height=5cm]{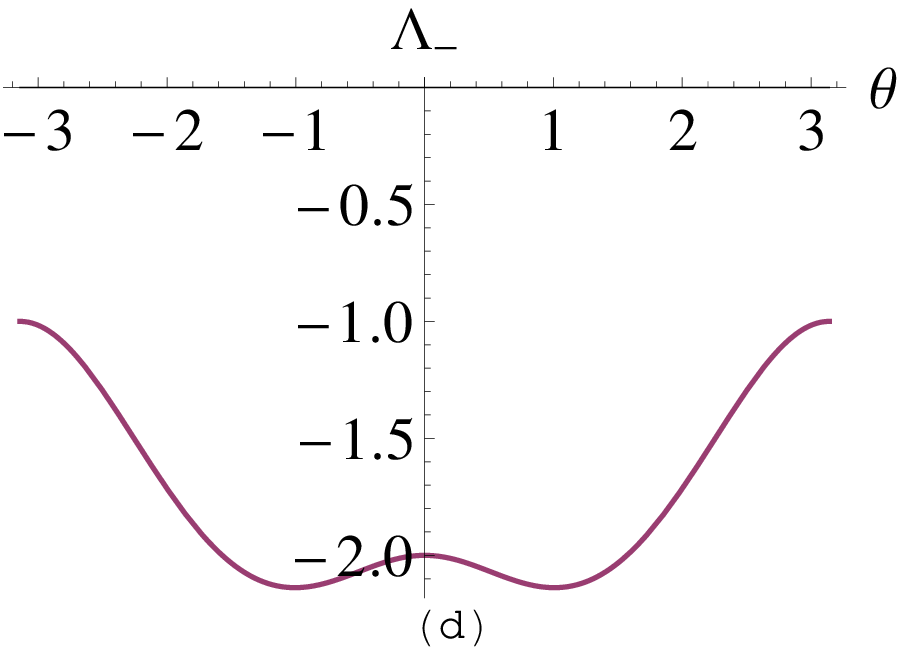}
\caption{(Color online) Representative plots of the real (thin black line) and imaginary
(thick line) parts of
$\Lambda_+$ and $\Lambda_-$ (calculated up to $O(q)$) vs $\theta$ for some chosen parameter
values $\lambda_2=1$, $\kappa_{\rho\rho}=2$, $\Delta\mu=1$ and $q=1$.
(a) $\Lambda_+$ vs $\theta$ for $\kappa_0\lambda_\rho=4$,
(b) $\Lambda_-$ vs $\theta$ for $\kappa_0\lambda_\rho=4$,
(c) $\Lambda_+$ vs $\theta$ for $\kappa_0\lambda_\rho=-4$, and
(d) $\Lambda_-$ vs $\theta$ for $\kappa_0\lambda_\rho=-4$.
In (a) the real part (thin line) is located in the region of negative
$\Lambda_+$ indicating stability
and the imaginary part (thick line) indicates the presence of propagating modes. In (c) and (d)
there are no real parts (black thin line) when $\kappa_0\lambda_\rho<0$.
Unlike the case for $l_s\rightarrow 0$, there are ranges of directions
in the plane where there are
only propagating modes at $O(q)$ (see text).}
\label{interfric}
\end{figure}

 The growth rate or relaxation rates of
the unstable and stable modes are also anisotropic and scale with
$\Delta\mu$. Representative plots of the real and imaginary parts of
$\Lambda_+,\Lambda_-$ as functions of $\theta$ for some chosen parameter
values are shown in Fig.~(\ref{interfric}) showing the presence of
propagating modes. The regions of instabilities and propagating modes are
clearly indicated. In particular, there are a few notable features as
displayed by Fig.~\ref{interfric}, consistent with the forms of the eigenvalues
(\ref{eigenweak}). For instance, for $\kappa_0\lambda_\rho<0$,
$\Lambda_+,\Lambda_-$ are wholly imaginary for all $\theta$ and unequal, i.e.,
the speed of the two modes are different in magnitude. In contrast, for
$\kappa_0\lambda_\rho>0$, the real parts vanish over an identical range of
$\theta$ for both the modes, that belongs to the shaded region in
Fig.~\ref{eigenint}, with unequal
imaginary parts, i.e., different speeds for the two modes. For the other values
of $\theta$, the real parts are nonzero with mutually opposite signs,
representing stable and unstable modes, with same speeds of propagation. The
overall differences with the eigenvalues for large (diverging) $\Gamma$ are
clearly visible.
Evidently, the model is overall stable at the linear order in $q$,
provided, $\lambda_\rho\kappa_0 \leq 0$. In this stable sector of the
parameter space, the results of Ref.~\cite{toner} that includes the
effects of the nonlinearities and noises should directly apply here.
Lastly, dry active matters are characterised by density segregation
in the steady states~\cite{bertin,ngo}. Our linearised treatment is unable to capture this.

 We now  consider briefly the case with $|\chi/\gamma_0-\bar\lambda|
\gg |\lambda_\rho\Delta\mu|$. Proceeding  as in
Sec.~\ref{highlin} above, the eigenvalues to the lowest order in $q$
and $\Delta\mu$ are given by
 \bea
 \Lambda = \pm\sqrt{{(\frac{\chi}{\gamma_0}-\bar\lambda)}\kappa_0\Delta\mu}
 q\sin\theta,
 \eea
 yielding instability for $(\chi/\gamma_0-\bar\lambda)\kappa_0\Delta\mu >0$ and oppositely
 moving propagating modes for $(\chi/\gamma_0-\bar\lambda)\kappa_0\Delta\mu <0$. These
 results are identical to the corresponding results in
 Sec.~\ref{highlin}.

\subsection{Weak friction limit}\label{weaklin}
We now analyse the linear instabilities for $\eta q^2 \gg\Gamma$.
 One of the eigenvalues $\Lambda$ of the linear stability matrix is non-zero
 at $O(q^0)$. We find
 \begin{equation}
 \Lambda=
 \frac{\zeta\Delta\mu}{2\eta}(\cos^2\theta-\sin^2\theta)\left[(\nu_1-1)\sin^2\theta
 - (\nu_1+1)\cos^2\theta\right].
 \end{equation}
  With a given choice for the sign of $\zeta\Delta\mu$ (say positive), $\Lambda > (<)0$
 for $\cos^2\theta-\sin^2\theta$ and $(\nu_1-1)\sin^2\theta-
 (\nu_1+1)\cos^2\theta$ having the same (opposite) signs
 and vice versa for $\zeta\Delta\mu <0$, suggesting
 instabilities for either sign of $\zeta\Delta\mu$. These results
 are identical to those in Ref.~\cite{sriram-aditi} for a bulk polar
 ordered active fluid. It is not a surprise that our results are
 same as those in Ref.~\cite{sriram-aditi}, for in the weak $\Gamma$
 limit, the active fluid layer in our model is effectively
 dynamically decoupled from the ambient fluid and hence acts as a
 free standing system, and hence, identical to the system considered
 in Ref.~\cite{sriram-aditi}.

\section{ Linear instabilities and measurements of $\Gamma$}\label{influgammaxx}

As our results above reveal, the magnitude of $\Gamma$ delineates
different regimes of the model. While  all these regimes display
generic long wavelength instabilities in the different regions of
the parameter space, the detailed nature of the instabilities and
the regions in the parameter space where they are present, vary
depending on $\Gamma$. For easy comparison, we provide here a
table (Table~\ref{table1}) which differentiates between the
instabilities in the three different regimes, as delineated by
$\Gamma$ (assume $|\chi/\gamma_0-\bar\lambda|\ll
|\lambda_\rho\Delta\mu|$):
\begin{table}
\begin{center}
\begin{tabular}{ |p{7cm}|p{6cm}|p{5cm}| }
\hline
High friction ($L\gg l_s$) & Intermediate friction ($L\ll l_s$) &
Weak friction [$L\ll (l_sd)^{1/2}$] \\
\hline Eigenvalues vanish at $O(q^0)$. & Again  eigenvalues vanish
at $O(q^0)$. & Nonzero eigenvalues: generic linear instabilities for
at $O(q^0)$ for both
signs of $\Delta\mu$; no propagating modes. \\
\hline $O(q)$: Generically linearly unstable at $O(q)$ for both
signs of $\Delta\mu$. In the plane at an angle $\theta_0$ (measured
with respect to the direction of the reference orientation) given by
$\tan^2\theta_0=(\nu_1+1)/(\nu_1-1)$ for $\lambda_\rho\kappa_0<0$,
or, for $\lambda_\rho\kappa_0>0$ and
$4\lambda_\rho\kappa_0\sin^2\theta_0<
(\lambda_2+\kappa_{\rho\rho})^2\cos^2\theta_0$: only propagating
modes at $O(q)$. Else instability even at $\theta=\theta_0$.
 &
 $O(q)$: Only propagating modes for $\kappa_0\lambda_\rho <0$. 
 Else, for
$\kappa_0\lambda_\rho
>0$, there are only propagating modes without any damping or growth
for angles $\theta$ in the plane satisfying $\cos^2\theta
(\kappa_{\rho\rho} +\lambda_2)^2 - 4\kappa_0\lambda_\rho\sin^2\theta
>0$ and instability elsewhere in
the plane.& -\\
\hline
\end{tabular}
\caption{Table describing the instabilities in different frictional
regimes.}
\label{table1}
\end{center}
\end{table}

   Despite the loose similarities between the nature of the
long wavelength instabilities for large and moderate $\Gamma$,
closer inspection reveals significant differences between the two
cases. With a large (formally diverging) $\Gamma$ $(l_s/L\ll 1)$,
the active fluid velocities $v_\alpha\sim O(q^0)$, where as, for
moderate $\Gamma$, $v_\alpha\sim O(q),\,\alpha=x,y$. Furthermore,
with a diverging $\Gamma$, the system is unstable in the full
parameter space to the lowest order in $q$ along all angles in the
polar plane, except for along the lines $\theta=\pm\theta_0,\pm
(\theta_0 +\pi)$. Along these special directions, there are only
propagating modes without any damping or growth (to the linear order
in $q$). At every other value of $\theta$, one mode is unstable.
Thus, for all (finite) values of the parameters and both signs of
$\Delta\mu$, there are moving instabilities with anisotropic speeds.
 In contrast, with intermediate interfacial friction ($l_s/L\gg 1$),
there are regions in the parameter space where there are only
propagating modes with no instabilities at the lowest order in $q$ for
any $\theta$; only in a subspace of
the parameter space, one encounters moving instabilities for either
sign of $\Delta\mu$. Even in such a parameter subspace, there are only
propagating modes at $O(q)$ for a range of $\theta$; for other values of
$\theta$, moving instabilities are present.

At this stage, it is useful to compare with available experimental results.
To do this, numerical estimates of the slip length $l_s$ or the slip
coefficient $\Gamma$ are needed. To our knowledge, systematic
measurements of $\Gamma$ or $l_s$ for active fluids are lacking.
Nonetheless, based on the available information we can make the
following comments. Ref.~\cite{arnabsaha} reports a hydrodynamic
length $\tilde l=(\eta/\Gamma)^{1/2}$ to be of the order of 10 $\mu
m$. Assuming $\eta\sim \eta' d$ and $d\sim 200
nm$~\cite{ewa} as the thickness of actin cortex, we find $l_s\sim
\tilde l^2/d\sim 10^3\mu m$, larger than the typical size of
cortical action layers ($\sim 100 \mu m$). Nevertheless, our
estimation of $l_s$ is not precise and hence it is difficult to
comment upon the experimental realisability of the high friction
case of {\em in-vitro} cortical actin layers in water based upon our estimates.
 On the
other hand, $L$ smaller than $l_s$ should correspond to the
intermediate friction case; we expect this to be realised in
experiments on cortical actin layers in water. Lastly, for $L< (l_s
d)^{1/2}\sim 10^{-1} \mu m$,  which is certainly small, the system
should behave as a free standing system. Equivalently, for a larger
system with $L>10^{-1} \mu m$, the dynamics of a free standing film
will be observed if the system is probed at length scales much
smaller than $ \sim 10^{-1}\mu m$; see, e.g., Ref.~\cite{arnabsaha}.
 The sensitive dependences of the long wavelength dynamics of the
model on $\Gamma$ may be used to make experimental estimates of
$\Gamma$ in a given system (i.e., for a fixed values of all other
parameters including $L$). We note that directly connecting our
theoretical predictions with experimental results is not an easy
task. Nonetheless, given the generic nature of our continuum active
fluid theories and noting that since an experiment is necessarily
performed on finite systems, long wavelength limit should imply
$2\pi/q \rightarrow L$, where $q$ is a wavevector of interest and
$L$ the system size, it is expected that all experiments that may be
described by the same long wavelength continuum equations should
display similar long wavelength linear instabilities, characterised
by their growth rates or thresholds of the linear instabilities. In
general, these properties should allow us to compare the theoretical
predictions with experiments, at least qualitatively, although
varying $L$ experimentally is expected to be a challenging task.
More specifically, we can make the following comments. First of
all, if linear instabilities are found to persist for all $\theta$,
then our above results indicate that $l_s/L\ll 1$ or a large
$\Gamma$: $\Gamma \gg \eta'/l_s$, hence, $\Gamma\gg \eta'/L$,
setting a lower bound for $\Gamma\sim \eta'/L$. On the other hand,
if linear instabilities are {\em not} found at $O(q)$ or found only
over a range of $\theta$, we can conclude $l_s/L\gg 1$, or, $\Gamma
\ll \eta'/L$, giving an upper bound on $\Gamma$. At the same time,
we must have $\eta q^2 << \Gamma,\eta\sim \eta' d$, yielding $\Gamma
\gg \eta' d/a^2$, where $a$ is a small scale ($\sim$ molecular cut
off), such that for $a^{-1}\lesssim q$, the continuum theory breaks
down. This provides a lower limit on $\Gamma$. In contrast, the
system behaves as a free standing 2D film for $L\ll
(\eta/\Gamma)^{1/2}$. Information on $\Gamma$ may also be obtained
by measuring the correlation functions of the local velocity fields,
orientation and density fluctuations and using the relations
(\ref{vxweak}) and (\ref{vyweak}) when the system is stable with an
intermediate $\Gamma$, i.e., with $\kappa_0\lambda_\rho <0$.
Velocity fields may be measured, e.g., by attaching a small bead
with the actin filament and tracking its instantaneous positions.
Orientation and density fluctuations may be measured by optical
methods and scattering experiments, respectively. Lastly, as we have
shown in Appendices (\ref{ambienthigh}), (\ref{ambientinter}) and
(\ref{ambientweak}) that the magnitude of bulk 3D fluid velocity
depends strongly on $\Gamma$ for $\eta'/L\gg \Gamma\gg \eta' d/a^2$.
Thus, measurement of the ambient 3D velocity field, e.g., by
tracking the position of a tracer particle, should also be helpful
in extracting numerical estimates on $\Gamma$.

\section{Nematic limit of the dynamics}\label{nemlim}
Until now we have considered polar active particles, so that the
corresponding dynamics is not invariant under $\bf p\rightarrow -p$.
In the nematic limit, the dynamics is invariant under $\bf
p\rightarrow -p$. Hence, active coefficients
$\lambda_2,\lambda_\rho,\kappa_0$ and $\kappa_{\rho\rho}$ and
equilibrium couplings $\chi$ and $\overline\lambda$ are zero. Thus,
to the lowest order, the dynamical equations in the strong friction
case are
 \bea
 \frac{\partial p_y}{\partial t}&=&-{1 \over
4\eta'q}[(\nu_1-1)q_y^2-(\nu_1+1)q_x^2]
\left[{\zeta \Delta\mu }\left(1- {2q_x^2 \over q^2}\right)p_y +
{\bar\zeta\Delta\mu q_xq_y \over  q^2}\rho\right],\label{pynemstrng}\\
 \frac{\partial\rho}{\partial t}&=&-\gamma_{\rho\rho}q^2 \rho +w\Delta\mu q_x^2\rho + w\rho_0\Delta\mu q_xq_yp_y.
 \eea
 As before, assume a time dependence of the form $\exp(\Lambda t)$ for the
 fluctuations. Then, to the lowest order in $q$
 \bea
 \Lambda =0,\frac{\zeta\Delta\mu
 q}{4\eta'}(\cos^2\theta-\sin^2\theta)\left[(\nu_1-1)\sin^2\theta-(\nu_1+1)\cos^2\theta\right].
 \eea
 Thus, with positive  $\zeta\Delta\mu$, $\Lambda$ is positive
 (negative) for $\cos^2\theta -\sin^2\theta$ and
 $(\nu_1-1)\sin^2\theta - (\nu_1+1)\cos^2\theta$ having the same
 (opposite) signs. Similarly for $\zeta\Delta\mu <0$.
 Thus, the system is unstable for both
 signs of $\Delta\mu$.

 For a finite $\Gamma$, to the lowest order in $\bf q$, the corresponding dynamical equations
 with nematic symmetry are
 \bea
 \frac{\partial p_y}{\partial t}&=&-\frac{Dq^2}{\gamma_0}p_y -
 \frac{\zeta\Delta\mu}{4\Gamma}\left[(\nu_1-1)q_y^2 -
 (\nu_1+1)q_x^2\right]\left(1-\frac{2q_x^2}{q^2}\right)p_y,\label{pynemweak}\\
\frac{\partial\rho}{\partial t}&=&-\gamma_{\rho\rho}q^2\rho
+w\Delta\mu q_x^2\rho + w\rho_0\Delta\mu
q_xq_yp_y.\label{rhonemweak}
 \eea
 Interestingly, Eqs.~(\ref{pynemweak}) and (\ref{rhonemweak}) are
 identical to those in Ref.~\cite{sriram-aditi-toner} for active
 nematics on a substrate. Thus, the results of
 Ref.~\cite{sriram-aditi-toner} are to hold here. We do not discuss
 these here in details.
Regardless of the details, in the nematic limit there are no
propagating waves and the instabilities are always static or
localised. In contrast, active polar ordered systems are characterised by the
presence of generic propagating modes and moving instabilities.
Finally, the eigenmodes in both the nematic and polar ordered
systems with strong interfacial friction with the embedding fluid
scale with $q$. However, for intermediate friction, the
eigenmodes for the nematic system scale as $q^2$, where as for the
corresponding polar ordered system, they scale as $q$.

\section{Summary}
\label{summ}

In this work, we have set up the generic coarse-grained dynamics of a thin layer of polar
ordered active particle suspensions frictionally coupled to the bulk isotropic
passive fluid with an arbitrary friction coefficient $\Gamma$.  In a linearised
treatment for small fluctuations around uniformly polar ordered states, we show
that our model describes a layer of wet active matter, dry active matter and a
free standing film, respectively, for $L\gg \eta'/\Gamma$,
$\eta'/\Gamma\gg L\gg (\eta' d/\Gamma)^{1/2}$ and $ \eta' d/\Gamma\gg L^2$.
The nature and the conditions for linear instabilities
in the long wavelength limit depend sensitively on $\Gamma$. These features may
be used to find estimates about $\Gamma$ in a given 2D active fluid layer
embedded in a bulk passive fluid.
We also discuss the nematic limit of
the dynamics and compare it with their polar analogues.

Our results
evidently highlight the crucial role played by the interfacial
friction and demonstrate how experimental knowledge about the linear
instabilities may be used to extract information about the friction
coefficient. Actual biological realisations of quasi-2D active fluids
have more complicated structures. Our work should be considered only
as a first step towards a more complete physical understanding of
such systems. We expect our results to be useful in understanding in-vitro
experiments on reconstituted layers of ordered actin filaments with
molecular motors in an embedding fluid (e.g., water). Experimental validation
of the $\Gamma$-dependences of the linear instabilities are expected to be
highly
challenging tasks. Nevertheless, we look forward to possible experimental
attempts to study the issues highlighted here. Lastly the formal similarities
between the dynamical equations with moderate interfacial friction
and those for a polar ordered system resting on a solid substrate
open up the possibilities of studying the physics of moderate
friction by performing experiments on an analogous system resting on
a solid substrate.


Our analyses are valid for small fluctuations around an ordered
state. Thus no conclusions may be drawn from our studies about the
eventual steady states in the event of the linearly unstable uniform
initial states. Numerical solutions of the full model equations
should yield valuable information in this regard. We made several  simplifying
assumptions while setting up our framework. For instance,  we have assumed the
active fluid layer to be inflexible and hence the
out-of-plane fluctuations are prohibited. However, this condition
may be violated for reconstituted actin filaments on a liposome.
Thus for better quantitative understanding of the experimental results, a thin
layer of active fluid with finite flexibility (i.e., with a finite
surface tension or bending modulus) should be studied. Secondly, the system may
not even be overall
flat and may have a finite curvature. In this case, our results should hold over
scales smaller than the radius of curvature. Our assumption of equal
friction coefficients on both the sides of the active system is also a
simplification. Generalisation to unequal frictions on both sides may be done in
a straightforward way.  It will
be interesting to study the diffusivity of a test particle inside
the 2D active polar system. It is well-known that the diffusivity of
a test particle in a free standing thin active fluid layer shows  starkly
unusual properties, e.g., dependences  on the thickness~\cite{bead}, in contrast
to the diffusivity of a small particle in a quasi-2D passive
fluid~\cite{delbruck}. Given our results here, it is expected that the
diffusivity of a test particle in a 2D polar ordered
medium is affected by the interplay of hydrodynamic interactions
by the embedding medium and the strength of the interfacial
friction.

 \section{Acknowledgement}
We would like to thank Jean-Francois Joanny, John Toner, Jacques Prost and Arnab Saha for
discussions at various stages of this work, and Frank J\"ulicher for
critical comments on the manuscript. NS would like to thank Krishanu Roy Chowdhury 
for helpful suggestions with plotting. One of the authors (AB) wishes
to thank the Max-Planck-Gesellschaft (Germany) and the Department of
Science and Technology/Indo-German Science and Technology Centre
(India) for partial financial support through the Partner Group
programme (2009).

\appendix
\section{Derivation of the full 2D generalised  Stokes equation for $v_i$}\label{stokesderi}

Here, we derive the full 2D generalised Stokes equation.
Using Eqs.~(\ref{totsigma}), (\ref{sigmatless})  in
Eq.~(\ref{forcebal}), the generalised Stokes Eq. may be written as
\bea
 \eta\nabla^2v_\beta + \Delta\mu\partial_\alpha(\zeta'(\rho) p_\alpha
p_\beta ) +{\nu_1 \over 2}\partial_\alpha(p_\alpha h_\beta +p_\beta h_\alpha )
-{\epsilon_0 \over 2}\partial_\alpha(p_\alpha
\partial_\beta\overline\mu_\rho +p_\beta\partial_\alpha\overline\mu_\rho )   +{1
\over 2}\partial_\alpha (p_\alpha h_\beta -p_\beta
h_\alpha)=\partial_\beta \Pi- F_\beta. \label{totfor} \eea

Using incompressibility $\Pi$ can be derived from (\ref{totfor}) as
\bea \Pi &=& \Delta\mu {\partial_\alpha\partial_\beta \over
\nabla^2}(\zeta'(\rho)(p_\alpha p_\beta ) +{\nu_1 \over
2}{\partial_\alpha\partial_\beta \over \nabla^2}(p_\alpha h_\beta
+p_\beta h_\alpha ) \nonumber \\
&&-{\epsilon_0 \over 2}{\partial_\alpha\partial_\beta \over
\nabla^2} (Ap_\alpha\partial_\beta\rho +Ap_\beta\partial_\alpha \rho +
\chi p_\alpha\partial_\beta{\boldsymbol\nabla}\cdot {\bf p}+\chi p_\beta
\partial_\alpha{\boldsymbol\nabla}\cdot {\bf p}) + {\partial_\beta F_\beta
\over \nabla^2}, \label{pifull} \eea 
where $1/\nabla^2$ is the
inverse  of $\nabla^2$.

Using this value of $\Pi$ in (\ref{totfor}), the Stoke's equation is derived as
\bea
\eta\nabla^2v_\beta &+& \Delta\mu P_{\beta\gamma}\partial_\alpha[\zeta' (\rho)
p_\alpha p_\gamma]+{\nu_1 \over 2}P_{\beta\gamma}\partial_\alpha
(p_\alpha h_\gamma +p_\gamma h_\alpha)-
\epsilon_0 P_{\beta\gamma}\partial_\alpha(Ap_\alpha\partial_\gamma\rho +
Ap_\gamma\partial_\alpha \rho + \chi p_\alpha\partial_\gamma{\boldsymbol\nabla}
\cdot {\bf p} + \chi p_\gamma\partial_\alpha{\boldsymbol\nabla}\cdot{\bf p})
\nonumber \\
&&+{1 \over 2}P_{\beta\gamma}\partial_\alpha(p_\alpha h_\gamma -p_\gamma h_\alpha)
=-P_{\gamma\beta}F_\gamma, \label{totstokes}
\eea
where $P_{\alpha\beta}$ is the transverse projection operator written as
$P_{\alpha\beta}=\delta_{\alpha\beta}-{\partial_\alpha\partial_\beta \over \nabla^2}$.
Let us redefine $P_{\gamma\beta}F_\gamma$ as $F_\beta$ given by (\ref{3dforce}).
Linearising about $p_x=1$, the Stokes equation (\ref{totstokes}) is simplified
to
\bea
\eta\nabla^2v_\beta &+& \Delta\mu P_{\beta x}\partial_\alpha[\zeta'p_\alpha]+
\Delta\mu P_{\beta\gamma}\partial_x[\zeta'p_\gamma]+{\nu_1 \over 2}P_{\beta x}
\partial_\alpha h_\alpha +{\nu_1 \over 2}P_{\beta\gamma}
\partial_xh_\gamma - {1 \over 2}P_{\beta x}\partial_\alpha h_\alpha +
{1 \over 2}P_{\beta\gamma}\partial_xh_\gamma \nonumber \\
&&-A{\epsilon_0 \over 2}P_{\beta\gamma}\partial_x\partial_\gamma\rho
-\chi{\epsilon_0 \over 2}P_{\beta\gamma}\partial_x\partial_\gamma
\nabla\cdot{\bf p} -A{\epsilon_0 \over 2}P_{\beta x}\partial_\alpha^2\rho
-\chi{\epsilon_0 \over 2}P_{\beta x}\partial_\alpha^2
\nabla\cdot{\bf p}=-F_\beta, \label{stokeslin}
\eea Now using Eq.~(\ref{free1}), we find $h_y=-{\partial\mathcal{F}
\over \partial p_y}= D\nabla^2p_y +\chi\partial_y\rho$. In addition,
$h_x$ acts as a Lagrange multiplier to enforce the constraint
$p^2=1$.  Notice that $h_y$ contributes terms which are higher order
in gradients in Eq.~(\ref{stokeslin}). Thus neglecting all the
higher order terms, the generalised Stokes equation up to the lowest
order in gradients is given by Eq.~(\ref{stokes}).

\section{ $F_\beta$ for high friction ($L\gg l_s$)}
\label{fbetahigh}

The velocity and hydrodynamic pressure for the subphase and
superphase are given by Eqs.~(\ref{vxup})-(\ref{pidown}). We impose
 incompressibility on the 3D ambient fluid: \bea
\partial_zv_z'=-\nabla_iv'_i\,\,\,\mbox{with i=x,y}, \label{incomp}
\eea for both $z>0$ and $z<0$. Fourier transforming  the in-plane
coordinates ${\bf x}=(x,y)$,
 \bea
\eta'(-q^2+\partial_z^2)v_z' &=& \partial_z\Pi', \\
\eta'(-q^2 +\partial_z^2)v_i' &=& iq_i\Pi', \,\,\, \mbox{and} \\
(-q^2+\partial_z^2)\Pi' &=& 0,
 \eea where $i=x,y$; ${\bf q}=(q_x,q_y)$ is the in-plane Fourier wavevector.
 The above equations can be
solved together to obtain the solutions for $v_x'$, $v_y'$, $v_z'$
and $\Pi'$. We write \bea
v_x' &=& (A_1 + B_1 z)\exp (-qz) \,\,\, \mbox{for $z>0$}, \label{vxup} \\
 &=& (A_2 +B_2z)\exp{(qz)} \,\,\, \mbox{for $z<0$}, \label{vxdown} \\
v_y' &=& (A_3 + B_3z)\exp{(-qz)} \,\,\, \mbox{for $z>0$}, \label{vyup} \\
&=& (A_4 +B_4z)\exp{(qz)} \,\,\, \mbox{for $z<0$}, \label{vydown} \\
v_z' &=& (C_1 + D_1z)\exp{(-qz)} \,\,\, \mbox{for $z>0$}, \label{vzup} \\
&=& (C_2 +D_2z)\exp{(qz)} \,\,\, \mbox{for $z<0$}, \,\,\, \mbox{and} \label{vzdown} \\
\Pi' &=& E_1\exp{(-qz)} \,\,\, \mbox{for $z>0$}, \label{piup} \\
&=& E_2\exp{(qz)} \,\,\, \mbox{for $z<0$}, \label{pidown} \eea where
coefficients $A_1, A_2,...,E_2$ are real or imaginary functions of
$\bf q$.

The incompressibility condition (\ref{incomp}) yields
 \bea
D_1 &=&-iq_xA_1-iq_yA_3+qC_1 = i{q_x \over q}B_1 + i{q_y \over q}B_3, \label{d1} \\
D_2 &=& -iq_xA_2 -iq_yA_4-qC_2 = -i{q_x \over q}B_2-i{q_y \over q}B_4. \label{d2}
\eea

The continuity of velocity or Eq.~(\ref{velcont}) gives
\bea
A_1=A_2 &=& v_x \label{vxa} \\
A_3=A_4 &=& v_y \label{vya} \\
C_1 &=& C_2 \label{cc}
\eea
As the active fluid film is two dimensional, there is no discontinuity over
the vertical gradient of $v_z'$ (since $v_z=0$). This allows us to write
\bea
\partial_zv_z'|_{z=\epsilon} =\partial_zv_z'|_{z=-\epsilon},
\eea
which yields using Eqs.~(\ref{vzup}) and (\ref{vzdown})
\bea
D_1=2qC_1 +D_2. \label{dd}
\eea

The tangential stress $F_i$ may be evaluated using the Stokes
equation (\ref{stokessubsup}). The Stokes equation for $v'_i$ yields
\bea \eta'\nabla_3^2\nabla_3\times v'_i=0. \label{stokesmod} \eea
Eq.~(\ref{stokesmod}) gives us further relations
 \bea
B_3 &=& -i{q_y \over q}D_1={q_y \over q_x}B_1 \,\,\, \mbox{and} \label{b1b3}\\
B_4 &=& i{q_y \over q}D_2 = {q_y \over q_x}B_2. \label{b2b4}
\eea

Using Eqs.~(\ref{d1})-(\ref{cc}), (\ref{dd}),
(\ref{b1b3}) and (\ref{b2b4}),
the $x$-component of 3D force $F$ is obtained as
\bea
F_x &=& \eta'(\partial_zv'_x+\partial_xv_z')|_{\epsilon}-\eta'(\partial_zv'_x
+\partial_xv_z')|_{-\epsilon} \nonumber \\
&=& \eta'[-2{q_x^2 \over q}v_x -2qv_x -2{q_xq_y \over q}v_y +i{q_x \over q}(D_1+D_2)
+B_1-B_2] \nonumber \\
&=& \eta'[-2{q_x^2 \over q}v_x-2qv_x-2{q_xq_y \over q}v_y] \nonumber \\
&=& -2\eta'qv_x, \label{fxfull} \eea where we have used
incompressibility of the 3D velocity in the last line. Similarly we
get the $y$-component of $F$ as \bea F_y=-2\eta'qv_y, \label{fyfull}
\eea 
using the no-slip condition equating $\bf v$ with the
in-plane components of $\bf v'$ at the active fluid layer.

\section{Form of $F_\beta$ for intermediate friction ($l_s\gg L\gg
(l_sd)^{1/2}$)}\label{intfric}
\label{fbetainter}

We start with
 \beq
 \eta'\frac{\partial v'_\alpha}{\partial z}|_{z=\epsilon} =
 \Gamma (v_\alpha'|_{z=\epsilon}-v_\alpha),\,\alpha=x,y.
 \eeq
 A similar condition exists at $z=-\epsilon$. Now using the forms of
 $v_x',v_y'$ and $v_z'$ as given by (\ref{vxup}), (\ref{vyup}) and
 (\ref{vzup}) we obtain
 \bea
 \eta'(-A_1 q + B_1)=\Gamma (A_1-v_x),\\
 \eta'(-A_3q + B_3)=\Gamma (A_3-v_y).
 \eea
 In the weak friction limit, $\Gamma\ll O(\eta'q)$ in the wavevector
 range of interest. Thus,
\bea
 \eta'(-A_1 q + B_1)=\Gamma (-v_x),\\
 \eta'(-A_3q + B_3)=\Gamma (-v_y).
 \eea
 or, equivalently,
 \bea
 \eta'\frac{\partial v_x'}{\partial z}=-\Gamma v_x,\\
 \eta'\frac{\partial v_y'}{\partial z}=-\Gamma v_y
 \eea
 at $z=\epsilon$. Similar considerations at $z=-\epsilon$ finally
 yields
 \bea
 \eta'[\frac{\partial v_\alpha'}{\partial
 z}|_{z=\epsilon}-\frac{\partial v_\alpha'}{\partial z}|_{z=-\epsilon}]=-2\Gamma
 v_\alpha.
 \eea
 This yields for the 2D generalised Stokes equation which $v_\alpha$
 satisfy
 \
 \bea \eta\nabla^2v_\alpha + \zeta\Delta\mu P_{\alpha\gamma}\partial_x p_\gamma + \zeta\Delta\mu
P_{\alpha x}\partial_yp_y +\bar\zeta\Delta\mu P_{\alpha
x}\partial_x\rho -2\Gamma v_\alpha=0, \label{stokesinter}.
\label{2dstokesweakgamm}
 \eea
 Now write Eq.~(\ref{2dstokesweakgamm}) in the Fourier space and
 neglect $\eta q^2 v_\alpha$ assuming $\eta q^2 \ll
 2\Gamma$. This yields Eqs.~(\ref{vxweak}) and (\ref{vyweak}).

\section{Velocity profiles of the ambient fluid}\label{3dv}

It is instructive to obtain the flow profiles of three-dimensional
velocity fields, that are created by the (small) fluctuations in
$\rho$ and $p_y$, in the three different regimes of our model
as delineated by the values of $\Gamma$.

\subsection{Large $\Gamma\; (L\gg l_s)$}
\label{ambienthigh}

In this case $v_i'(x,y,z=\pm\epsilon)=v_i(x,y),\,i=x,y$. Since
$v_z'(z=\pm\epsilon)=0$, from (\ref{vzup}) and (\ref{vzdown}),
$C_1=0=C_2$. Using the no-slip condition on $v_i'(z=\pm)$ and the
3D incompressibility of $v_\alpha',\alpha=x,y,z$, $(iq_x v'_x + iq_y)|_{z=\pm\epsilon}
=0=\frac{\partial v_z'}{\partial z}|_{z=\pm}$ in the Fourier space.
This yields $D_1=0=D_2$. Thus, $v_z'=0$ everywhere above and below
the active fluid layer. Hence, the flow in the surrounding fluid is
actually 2D, parallel to the active fluid layer. We further find
$B_1=0=B_2$ and $B_3=0=B_4$. Thus in the Fourier space,
\bea
v_i'(q_x,q_y,z)&=&v_i(q_x,q_y)\exp(-qz),\,z>0,\nonumber \\
&=& v_i(q_x,q_y)\exp(qz),\,z<0.
\eea
Therefore, $v_i'$ has the same form as $v_i$ with an exponentially
damped amplitude by a factor $\exp(-q|z|)$ and hence shows the same instabilities at $O(q)$.

\subsection{Intermediate $\Gamma\; (l_s\gg L\gg (l_s d)^{1/2})$}
\label{ambientinter}

In the intermediate friction case, the 3D shear stress balance is given by
\bea
\eta'{\partial v_i' \over \partial z}|_{\epsilon} &=&-\Gamma v_i|_{\epsilon}, \label{interup} \\
\eta'{\partial v_i' \over \partial z}|_{-\epsilon}&=& \Gamma v_i|_{-\epsilon}. \label{interdown}
\eea
Using the above equations and (\ref{vxup}), (\ref{vxdown}), (\ref{vyup}) and
(\ref{vydown}), we get a set of relations between the couplings given by
\bea
-qA_1+B_1 &=& -{\Gamma \over \eta'}v_x, \label{rel1} \\
qA_2+B_2 &=& {\Gamma \over \eta'}v_x, \label{rel2} \\
-qA_3+B_3 &=& -{\Gamma \over \eta'}v_y, \label{rel3}\\
qA_4+B_4 &=& {\Gamma \over \eta'}v_y. \label{rel4}
\eea

Now using (\ref{rel1}), (\ref{rel2}), (\ref{rel3}), (\ref{rel4}), 3D incompressibility
($\nabla\cdot {\bf v}'=0$) of ambient fluid, 2D incompressibility ($\nabla_\bot\cdot
{\bf v}=0$) of active fluid layer and equations (\ref{b1b3}) and (\ref{b2b4}),
we can show that
\bea
B_1=B_2=B_3=B_4 &=& 0 \,\,\,\,\mbox{and}, \label{b0} \\
D_1=D_2 &=& 0. \label{d0}
\eea

Hence we obtain, $v_z'=0$ for all $z$ identically and $v_i'(q_x,q_y,z)=\Gamma v_i
\exp(-q|z|)/(\eta' q)$ and has the same dependences on $\rho$ and
$p_y$ as for large $\Gamma$. Thus, $v_\alpha',\,\alpha=x,y,z$ is again 2D.
Nonetheless, $v_i'$ for moderate $\Gamma$ is different from $v_i'$ for large
$\Gamma$, since the solutions for $\rho$ and $p_y$ have very different
explicit forms for moderate $\Gamma$. In particular, $v_i'(x,y,z)$ shows instability
only if $\kappa_0\lambda_\rho>0$, the same condition for instability for the 2D active fluid layer. Notice that in the present case in addition to the $\exp(-q|z|)$ factors, $v_i'$ is further scaled down in comparison
with $v_i$ by a factor $\Gamma/(\eta'q)$, and hence should be small in magnitude.

\subsection{Small $\Gamma\; (L^2 \ll l_s d)$}
\label{ambientweak}

In the limit of very small $\Gamma$, we have $\eta'\frac{\partial v_i}
{\partial z}|_{z=\pm}\approx 0$. i.e., we effectively have the stress-free boundary
condition on $v_i'$ at $z=0$. In addition, $v_z'=0$ at $z=0$. Now using
the results in Sec.~\ref{intfric}, we find $A_1=A_2=B_1=B_2=A_3=A_4=B_3=B_4=0$,
i.e., $v_i'=0$ at all $z>0$ and $z<0$ identically. In addition, $v_z'=0$
everywhere. Thus, the 3D velocity field vanishes.


\begin{thebibliography}{99}
\bibitem{cortex} The thin active fluid layer in our model corresponds
to the cortical actin layer in a cell, where as the embedding fluid
in our model represents the bulk cytoplasm. While these straight forward
relations between our model and the cell cortex seem self-evident and simple,
it should be remembered that in an eukaryotic cell, there are no sharp
bloundaries between the cortical actin layer and the bulk cytoplasm. In
addition, the bulk cytoplasm is not a passive fluid; it has active processes going on in it.
\end{thebibliography}
\end{document}